\documentclass[preprint]{aastex}

\shortauthors{Palma, Majewski, \& Johnston}
\shorttitle{Orbital Poles of Milky Way Satellites}

\begin{document}

\title{On the Distribution of Orbital Poles of Milky Way Satellites}
\author{Christopher Palma\altaffilmark{1}, Steven R. Majewski\altaffilmark{2}}
\affil{University of Virginia, Department of Astronomy \\
       PO Box 3818, University Station, Charlottesville, VA  22903\\
       Email:  cp4v@virginia.edu, srm4n@virginia.edu}
\authoraddr{P. O. Box 3818, Charlottesville, VA  22903-0818}
\altaffiltext{1}{Current address:  Department of Astronomy \& Astrophysics,
Penn State University \\ 525 Davey Laboratory, University Park, PA  16802}
\altaffiltext{2}{David and Lucile Packard Foundation Fellow; Cottrell Scholar
of the Research Corporation; National Science Foundation CAREER Fellow;
Visiting Associate, The Observatories of the Carnegie Institution of Washington}

\and

\author{Kathryn V. Johnston}
\affil{Van Vleck Observatory, Wesleyan University \\
       Middletown, CT, 06459 \\
       Email:  kvj@astro.wesleyan.edu}
\authoraddr{Astronomy Department, Wesleyan University, Middletown, CT}

\begin{abstract}

In numerous studies of the outer Galactic halo some evidence for
accretion has been found.  If the outer halo did form in part or wholly
through merger events, we might expect to find coherent streams of
stars and globular clusters following similar orbits as their parent
objects, which are assumed to be present or former Milky Way dwarf
satellite galaxies.  We present a study of this phenomenon by assessing
the likelihood of potential descendant ``dynamical families'' in the
outer halo.  We conduct two analyses:  one that involves a statistical
analysis of the spatial distribution of all known Galactic dwarf
satellite galaxies (DSGs) and globular clusters, and a second, more
specific analysis of those globular clusters and DSGs for which full
phase space dynamical data exist.  In both cases our methodology is
appropriate only to members of descendant dynamical families that
retain nearly aligned orbital poles today.  Since the Sagittarius dwarf
(Sgr) is considered a paradigm for the type of merger/tidal interaction
event for which we are searching, we also undertake a case study of the
Sgr system and identify several globular clusters that may be members
of its extended dynamical family.

In our first analysis, the distribution of {\em possible} orbital poles
for the entire sample of outer ($R_{gc} > 8$ kpc) halo globular
clusters is tested for statistically significant associations among
globular clusters and DSGs.  Our methodology for identifying possible
associations is similar to that used by \citet{lb295} but we put the
associations on a more statistical foundation.  Moreover, we study the
degree of possible dynamical clustering among various interesting {\em
ensembles} of globular clusters and satellite galaxies.  Among the
ensembles studied, we find the globular cluster subpopulation with the
highest statistical likelihood of association with one or more of the
Galactic DSGs to be the distant, outer halo ($R_{gc} > 25$ kpc), second
parameter globular clusters.  The results of our orbital pole analysis
are supported by the Great Circle Cell Count methodology of
\citet{kvj96}.  The space motions of the clusters Pal~4, NGC~6229,
NGC~7006, and Pyxis are predicted to be among those most likely to show
the clusters to be following stream orbits, since these clusters are
responsible for the majority of the statistical significance of the
association between outer halo, second parameter globular clusters and
the Milky Way DSGs.

In our second analysis, we study the orbits of the 41 globular clusters
and 6 Milky-Way bound DSGs having measured proper motions to look for
objects with both coplanar orbits and similar angular momenta.
Unfortunately, the majority of globular clusters with measured proper
motions are inner halo clusters that are less likely to retain memory
of their original orbit.  Although four potential globular cluster/DSG
associations are found, we believe three of these associations
involving inner halo clusters to be coincidental.  While the present
sample of objects with complete dynamical data is small and does not
include many of the globular clusters that are more likely to have been
captured by the Milky Way, the methodology we adopt will become
increasingly powerful as more proper motions are measured for distant
Galactic satellites and globular clusters, and especially as results
from the Space Interferometry Mission (SIM) become available.

\end{abstract}


\section{\bf Introduction}

Models for structure formation in the universe that include a dominant
cold dark matter (CDM) component predict a hierarchical formation
process where large structures are formed by the merging of smaller CDM
halos.  Numerical simulations \citep[e.g.,][]{moore, kly99} seem to
indicate that galaxies form in a similar fashion as do galaxy clusters,
as subgalactic CDM halos merge to form galaxy sized halos.  These
simulations support the idea that the Milky Way formed as an
aggregation of smaller units, however, there is some controversy
because the simuluations overpredict the number of subgalactic halos
that remain at $z=0$ in a Local Group type environment.  In the Local
Group, there are two populations of subGalactic objects that may be
related to the CDM halos in numerical simulations:  globular clusters
and dwarf galaxies.  It is possible that there remains information on
the growth of structure in the Milky Way system encoded in the current
globular cluster and dwarf satellite galaxy (hereafter, DSG)
populations found in orbit around the Milky Way.

Globular clusters are often used as archetypal objects, and their
current properties can not only be used to constrain their formation
and evolutionary histories, but those of the Galactic stellar
populations they trace.  In the Milky Way, the population of globular
clusters has traditionally been split into an inner and an outer population,
separable by metallicity \citep{zinn80}.  More recently, it has been shown
that these ``halo'' (outer) and ``disk'' (inner) populations can be
further subdivided when additional properties are considered in
addition to metallicity.

The properties of the outer ($R_{gc} > 8$ kpc) globular clusters of the
Milky Way suggest that they are, and trace, a distinct population with
formation and evolutionary histories different from those of the more
tightly bound, inner ($R_{gc} < 8$ kpc) globular clusters.  The outer
globular clusters share similar kinematical, metallicity, age, and
spatial distributions as halo stars \citep[e.g.,][]{zinn85, zinn96}
and are thus usually assumed to be representative of the halo stellar
population.  It is more difficult to assign individual inner globular
clusters to specific stellar populations, because of the overlap in
properties between the bulge, thin disk, thick disk, and halo
populations near the center of our Galaxy; however, for the most part
the inner globular clusters tend to have kinematical, metallicity, and
spatial distributions closer to those of the bulge or disk than of the
halo \citep[e.g.,][]{zinn85, az88, minn95, zinn96},
although \citet{burk97} use dynamical arguments to assign some of the
highest metallicity, inner globular clusters to an inner halo
population, distinct from a ``bar'' population and a 5 kpc ring
population.  The kinematical and spatial differences found between the
inner and outer globular cluster subpopulations (specifically the
existence of {\em retrograde} orbiting globular clusters found among
the latter group) support the Galactic formation scenarios of
\citet{searle77} and \citet{SZ}, who proposed that the outer halo of the
Milky Way may have formed through the infall and accretion of gas and
stars from ``fragments'' after the collapse of the proto-Galactic cloud
that produced the inner Milky Way.  The accretion into the halo of
globular clusters that formed in fragments can account for the observed
apparent age spread in globular clusters that may be as large as
$\sim$5 Gyr by some accounts \citep[for a recent review see][]{ata}.
Although the magnitude of the age spread among globular clusters is
still uncertain \citep[e.g.,][]{stet, vdb} any significant halo
age spread ($> 1$ Gyr) is incompatible with the timescale for a single
collapse for halo formation as originally proposed in the \citet{els}
model.

Studies by \citet{wekdem76} and \citet{lb82} found spatial alignments
among the DSGs, fueling speculation that these objects may be the
fragments proposed in the \citet{SZ} accretion model of the Galactic
halo.  \citet{wekdem76} postulated that a group of six DSGs and four
red horizontal branch globular clusters, which they denoted the
``Magellanic Plane Group'', are relics of a past tidal interaction
between the Magellanic Clouds and the Galaxy since the DSGs, clusters,
and the Magellanic Clouds lie near a great circle that is nearly
coincident with the Magellanic (HI gas) Stream.  Subsequently,
\citet{wek79}, using contemporary radial velocity data, presented
evidence for motion along a single orbit by the Magellanic Plane DSGs
and globular clusters, which provided further support for the tidal
disruption hypothesis.  \citet{lb82}, adopting a different orbital
indicator, suggested that one could identify objects on similar orbits
(i.e.  remnants of a single merger event) by looking at their angular
momentum axes, assumed to be given by $\vec{r}\times\vec{p}$ where
$\vec{r}$ is the Galactocentric radius vector of an object and
$\vec{p}$ is the position angle of the tidal extension of the object.
For example, he noted a coincidence between the tidal elongation of the
Ursa Minor dwarf galaxy and the orientation of the Magellanic Stream.
Looking at the spatial distribution of all known Milky Way DSGs and
their tidal elongations, Lynden-Bell identified two ``streams'' of
objects, a Magellanic stream\footnote[3]{Hereafter, when we refer to
the ``Magellanic stream'', we mean the great circle defined by
\citet{lb82} that contains the LMC, SMC, Ursa Minor, and Draco DSGs.
This is to be contrasted to the ``Magellanic Stream'', the large
complex of HI gas associated with the Magellanic Clouds.}, and an
``FLS'' stream that contains the Fornax, Leo I, Leo II, and Sculptor
dwarf galaxies.  According to \cite{lb82}, these spatial alignments may
have arisen from the tidal disruption of a Greater Magellanic Galaxy
and a Greater Fornax Galaxy.  Given recent evidence for the disruption
of DSGs themselves \citep[e.g., Carina and Ursa Minor;][]{haloI,
palma01}, it is possible that these dwarf galaxies represent an
intermediate phase in the total accretion of larger, LMC-like
satellites by the Milky Way.

Both \citet{wekdem76} and \citet{lb82} included specific
globular clusters, generally those in the outer halo, in their
alignment schemes.  Interestingly, it is these same clusters that
played a significant role in shaping the original \citet{SZ}
picture and which have continued to spark interest in the possibility
that the Milky Way halo continues to assimilate debris from the
disruption of chemically distinct systems.  In an influential recent
study, \citet{zinn93a} found evidence for significant kinematical
differences between two populations of halo globular clusters
discriminated by a combination of the \citet{lee94} index of
horizontal branch morphology and [Fe/H].  In this new scheme, \citet{zinn93a}
refers to the two subdivisions of outer halo globular clusters
as the ``Old Halo'' and ``Young Halo'' globular clusters under the
assumption that the second parameter of horizontal branch morphology is
age.  The ``Young Halo'' globulars are found to have a mean rotational
velocity that is retrograde and with a large line-of-sight velocity
dispersion of $\sigma_{LOS} = 149\pm24$ km/sec.  This is in contrast to
the ``Old Halo'' globular clusters, which have a mean prograde
rotational velocity and a much smaller $\sigma_{LOS}$.  \citet{zinn93a}
suggests that the observed flattening in the spatial distribution of
the combined Old Halo and Disk globular populations, as well as the
correlation of [Fe/H] to Galactocentric radius ($R_{gc}$) within those
combined populations, imply that the Old Halo globulars and the Disk
globular clusters together may be products of the same formation
mechanism, perhaps an ELS-like collapse.  On the other hand, the
spherical spatial distribution, lack of a metallicity trend with
$R_{gc}$, and the possible retrograde rotation of the Young Halo
globular clusters implies that they may have formed separately than the
Disk$+$Old Halo globular clusters, and then were later accreted by the
Milky Way \'{a} la \citet{SZ}.  In a more recent study, \citet{zinn96}
subdivides the halo globular clusters further, into three
populations, and adopts terminology that is less specific regarding the
possible origin of the second parameter effect.  The ``RHB'' (red
horizontal branch) group is essentially the same as the Young Halo
group from \citet{zinn93a}.  However, the Old Halo group he now splits into
a ``MP'' (metal-poor) group with $[$Fe/H$] < -1.8$ and a ``BHB'' (blue
horizontal branch) group with $-1.8 < [$Fe/H$] < -0.8$ (the metallicity
range where the second parameter effect operates).  We adopt the more
recent terminology of RHB vs. BHB/MP since it makes no assumption as
to the origin of the second parameter effect.

It is now recognized that many of the most distant outer globular
clusters are predominantly of the RHB type, and \citet{srm94} shows
that the outer halo globular cluster/DSG connection
may pertain to the origin of the second parameter effect.  Like
\citet{wekdem76}, who included several red horizontal branch clusters
as part of their Magellanic Plane group, \citet{srm94} found a spatial
alignment between a sample of Young Halo globular clusters and the FLS
stream galaxies.  Majewski found that if one fits an orbital plane to
the positions of the FLS stream DSGs similar to Lynden-Bell's (1982)
orbital plane for the FLS stream, the positions of the most distant
outer halo, red horizontal branch (young) globular clusters (as well as
the more recently discovered Sextans dwarf and the Phoenix dwarf) are
found to be correlated with the best-fit plane.  \citet{fusi} also
fit a plane to the spatial distribution of the Galactic DSGs, and
found that many of the globular clusters considered to be younger than
the majority of Galactic globular clusters (which are all RHB type) lie
on their best fit plane.

The most striking evidence for a tidal capture origin for some outer
halo globular clusters comes from the recently discovered (Ibata et
al.\ 1994, 1995) Sagittarius dwarf spheroidal (Sgr).  This DSG is
currently $\sim$16 kpc from the Galactic center, closer than any of the
other Milky Way DSGs.  A consequence of its apparently
small perigalacticon is that the Sagittarius dwarf shows evidence for
ongoing tidal disruption by the Milky Way \citep{ib95, mmetal98, kvj99}.
Of particular interest to our
discussion here are four globular clusters (M54, Arp~2, Ter~7,
and Ter~8) with Galactocentric positions and radial velocities very
similar to those of Sagittarius.  It appears that at least some of
these four globular clusters originally belonged to Sagittarius and are
in the process of being stripped from their host by the Milky Way.
Recently, \citet{dd00} have determined the proper motion of Pal~12 and
argue that it too may have originally belonged to the Sgr.

Lynden-Bell \& Lynden-Bell (1995, hereafter LB$^{2}$95) recently
pioneered a technique for identifying other potential cluster/DSG
associations using the positions and radial velocities of the entire
sample of globular clusters and DSGs.  They identify
candidate streams similar to the Magellanic stream and FLS stream
by selecting families of objects whose ``polar paths'' (great circles
identifying all possible locations of their orbital poles, see \S3.1 and
\S5 below) share a nearly common intersection point and that also have
similar orbital energies and angular momenta as derived using current
radial velocities and an assumed Milky Way potential.  We revisit
the technique of LB$^{2}$95 in our attempt to address the following
questions:

\begin{itemize}

\item Can we improve the case for association of globular clusters with
DSGs?  Can we develop a more
statistical foundation for this suggestion?

\item Can we point to specific dynamical families in the halo to make
previously proposed associations of DSGs and/or globular
clusters less anecdotal?

\item Can we provide specific targets for follow-up study to test the
``dynamical family'' hypothesis?

\item Can we verify the suggestion that it is the RHB globular clusters
that are more associated with the tidal disruption process?

\item  What are the limitations in this type of analysis?

\end{itemize}

Previous, related investigations have all relied solely on positional
alignments (and in some cases radial velocities), or have estimated
orbital properties relying on assumptions about the shape of the
Galactic potential and the objects' transverse velocities.  In this
work, we first reinvestigate positional globular cluster and DSG
alignments by applying statistical tests to the results of an
LB$^{2}$95 type ``polar path'' analysis.  We then search for possible
dynamical groups among the (still relatively small sample of) Galactic
globular clusters and DSGs having available radial velocity {\em and}
proper motion measurements.

\section{The Sample of Halo Objects}

Positions and radial velocities of all globular clusters were adopted
from the 22 June 1999 World Wide Web version of the compilation by
\citet{mwgc}.  Satellite galaxy positions were taken from the
NASA/IPAC Extragalactic Database (NED\footnote[4] {The NASA/IPAC
Extragalactic Database (NED) is operated by the Jet Propulsion
Laboratory, California Institute of Technology, under contract with the
National Aeronautics and Space Administration.}), radial velocity data
from the recent review by \citet{mm98} and proper motions (for all
objects) from various sources, which are listed in Table \ref{pmtab}.  In cases
where multiple proper motions have been published for globular clusters
and DSGs, we have selected the measurement with the smallest random
errors. The proper motion that was adopted for analysis
is listed first in Table \ref{pmtab} for objects with multiple measurements.
However, for most globular clusters with multiple independent proper
motion measurements, the position of the orbital pole is fairly
insensitive to the differences between measurements.  Exceptions are
discussed in \S6.1.

For the DSGs and globular clusters in our analysis, we used the
adopted position, distance, radial velocity, and proper motion to
determine Galactic space velocities, $(U,V,W)$, and their one sigma
errors using the formulae from \citet{js87}.  These
velocities were transformed to the Galactic standard of rest using the
basic $(U,V,W)$ solar motion \citep{galast} of
$(-10.4,+14.8,+7.3)$ km/sec (the difference between this value and any
of the more recent determinations is significantly below the proper
motion velocity errors) and a rotational velocity of the Local Standard
of Rest (LSR) of $\theta_{0} = 220$ km/sec \citep{klb86}.
The Galactocentric Cartesian radius vectors, $(X,Y,Z)$, for the
DSGs and clusters were calculated using an adopted value of 8.5
kpc \citep{klb86} for the solar Galactocentric radius.  In instances
where orbital parameters were calculated for our sample objects,
the Galactic potential was assumed to be that of \citet{kvj95}

While proper motion errors tend to be large (sometimes of order 100\%)
for most objects, the radial velocity errors are often very small
compared to the magnitude of the radial velocity itself.  The
propagated errors in the $(U,V,W)$ velocities depend strongly on the
ratio of the magnitude of the radial velocity to the magnitude of the
tangential velocity, after one transforms to the Galactic Cartesian
system.  For example, Ursa Minor has a radial velocity of $-248 \pm 2$
km/sec, while its proper motion translates to a transverse velocity
of magnitude $30 \pm 40$ km/sec.
Even though its proper motion error is large, since its radial velocity
makes up the majority of its total space velocity, the error in its
total velocity is only $\sim 20$\%.  This ``reducing'' effect in the total
error is reflected in the length of the Arc Segment Pole Families (see
\S 3) for some of those objects with large proper motions errors (e.g., Ursa
Minor).

The final sample used in \S5 includes the 147 Milky Way globular
clusters from the \citet{mwgc} compilation and the LMC, SMC, Draco,
Ursa Minor, Sculptor, Sagittarius, Fornax, Leo I, Leo II, Sextans, and
Carina of the Milky Way DSGs.  We also included the Phoenix
dwarf galaxy, however, this object was left out of some of our analyses due to
its uncertain connection with the Milky Way.  From this sample of 147
globular clusters and 12 galaxies, we collected proper motion data from
the literature for 41 clusters and 6 galaxies (Table \ref{pmtab}), which we use
for the orbital pole analysis in \S6.

\section{The Orbital Pole Family Technique}

If we assume that the outer halo of the Milky Way was formed at least
partially through the tidal disruption of dwarf galaxy-sized objects,
we may expect to observe the daughter products of these mergers.  As
described in the Introduction, there are spatial alignments of 
DSGs and young globular clusters that suggest they may be sibling
remnants of past accretion events.  We wish to improve upon previous
searches for spatial alignments by identifying groups of these objects
that share a nearly common angular momentum vector.  The LB$^{2}$95
technique relies fundamentally on positional data for these clusters.
These data are not going to change in any significant way, and the only
way to improve on this basic technique is through more complete,
statistical analyses of the sample.  We do this here.  However, by
taking advantage of proper motion data, an improved, modified
LB$^{2}$95 technique can be applied, and this new approach will always
increase in usefulness as more and better proper motion data become
available.  Thus, our search technique has roots in the methodology of
LB$^{2}$95, but we take two steps forward from their analysis:  (1)  We
take advantage of a simple two-point angular correlation function
analysis to address possible alignments in a stellar populations
context, and (2) we take advantage of the growing database of orbital
data for outer halo objects to search for associations among refined
locations for the LB$^{2}$95 ``polar paths''.

\subsection{Constructing an Orbital Pole Family}

\begin{figure}
\plotone{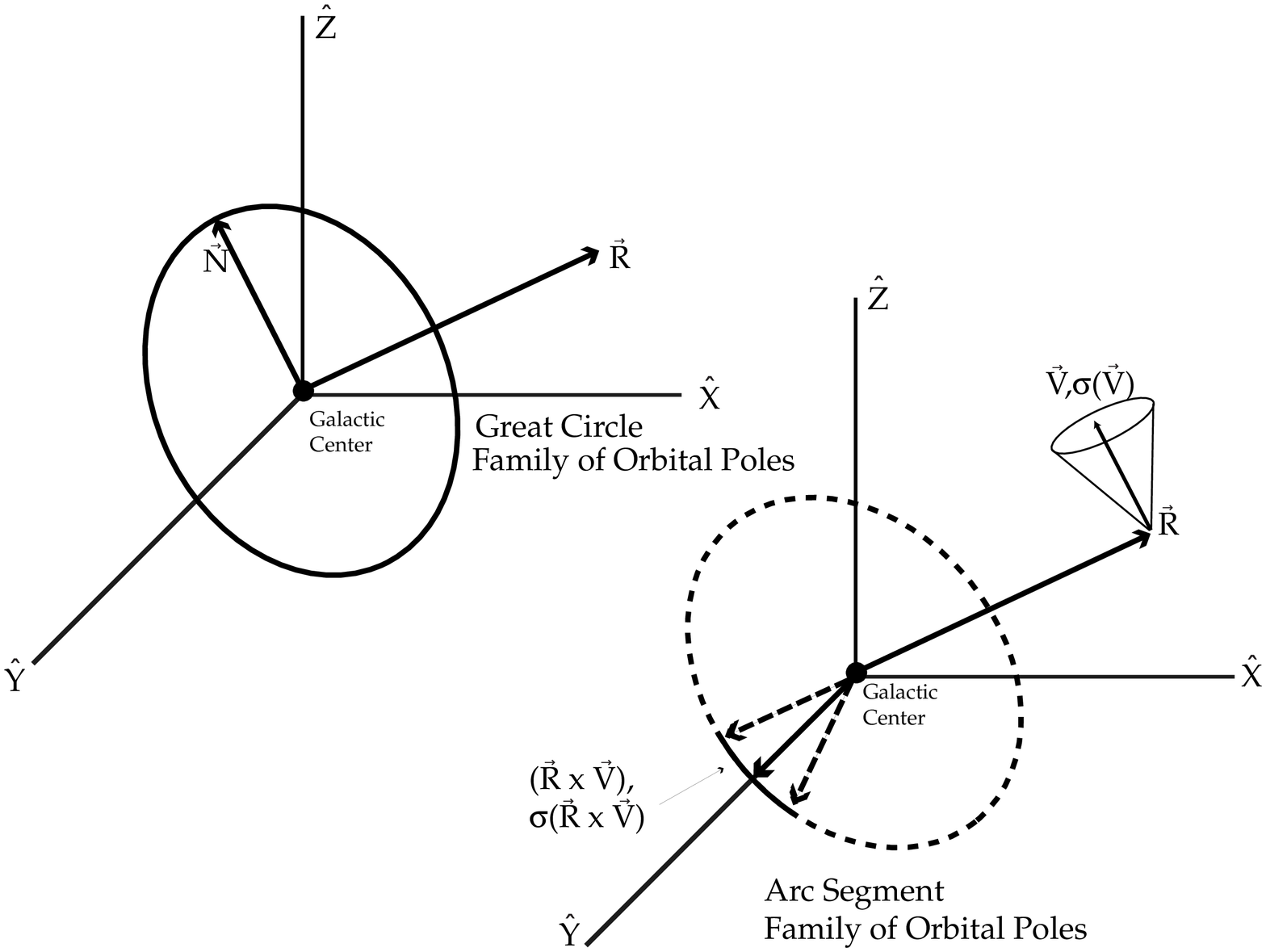}
\caption{On the upper left, the geometry of
orbital pole families for a Milky Way satellite, as per
the method of Lynden-Bell \& Lynden-Bell (1995).  Any object that has a
radius vector $\vec{R}$, will have a family of possible normal vectors, 
$\vec{N}$.  In the upper part of this figure, the vector labeled $\vec{N}$ is
only one of the possible normals, and the circle defines the positions of the
endpoints of all possible vectors, $\vec{N}$.  The lower right figure
depicts the construction of an arc segment pole family, in our
improved methodology using proper motion data.  The true pole is
defined by $\vec{R} \times \vec{V}$ where $\vec{V}$ is the object's space
velocity.  If one includes the 1-$\sigma$ error in the space velocity, the
resulting velocity and error cone limits the possible poles to lie along an
arc segment on the great circle.
\label{fig:polegeom}}
\end{figure}

As discussed by LB$^{2}$95, from knowledge of only its $\vec{R_{gc}}$
vector one can construct a family of possible orbital poles for each
Galactic satellite.  The basic assumption underlying pole family
construction is that each satellite orbits in a plane containing the
object's current position and the Galactic center.  The family of
possible orbital poles is simply the set of all possible normals to the
Galactocentric radius vector for a particular satellite (Figure
\ref{fig:polegeom}).  Clearly it is desirable to limit further the
family of possible orbital poles if at all possible.  There are two
limiting cases:  If one has no information save the object's Galactic
coordinates, the family of possible normals traces out a great circle
on the sky.  This is the basis of the LB$^{2}$95 method:  Any set of
objects constituting a {\em dynamical family} (i.e., a group of objects
from a common origin and maintaining a common orbit), no matter how
spread out on the sky, will have ``Great Circle Pole Families'' (GCPFs)
that intersect at the same pair of antipodal points on the celestial
sphere.  Thus, searching for possible dynamical families, orbiting in
common debris streams, means plumbing the set of all GCPFs for common
intersection points.  As a further constraint, LB$^{2}$95 derived
``radial energies'' (an approximation to the orbital energy that are derived
using measured heliocentric radial velocities) to eliminate objects
from streams with grossly different orbital parameters.

The other limiting case occurs if one knows the space velocity for the
object with infinite precision.
Then, the pole family consists of one point, the orbit's
true pole.  In reality, however, space motions of Galactic satellites
have fairly large uncertainties, generally, in large part due to the
proper motion errors.  Thus, we never truly achieve a well-defined
orbital pole point.  However, even with rough space motions we can
constrain the true pole to lie along an arc segment (an ``Arc Segment
Pole Family'', or ASPF) rather than a great circle (Figure
\ref{fig:polegeom}).  The better the space motion errors, the smaller
the arc segment, which, in the limit of no error, is a point.
LB$^{2}$95 provided lists of potential streams derived with their
GCPF$+$radial energy technique with the understanding that in the end,
proper motions must be measured to confirm stream membership.  Although
proper motions remain unavailable for the majority of the objects found
in the streams of LB$^{2}$95, we can search the sample of globular
clusters and DSGs with proper motions for associations of
their ASPFs, which produces better defined streams than those selected
with the GCPF method.

\subsection{Physical Limitations of the GCPF Technique}

Although techniques exist \citep[e.g., the Great Circle Cell Counts technique
of][]{kvj96} to search for the stellar component of tidal debris
among large samples of Galactic stars, orbital pole families can be a
powerful tool in the search for remnants of past merger events among
even small samples of objects in
dynamical associations spread out over the celestial
sphere.  In a spherical potential, the daughter products of a tidal
disruption event should have nearly identical orbital poles, indicative
of a common direction of angular momentum.  In such an ideal case, the
GCPFs of the daughter objects will all intersect at a pair of antipodal
points on the sky (another way of denoting that the objects all lie in
one plane).

\begin{figure}
\plotone{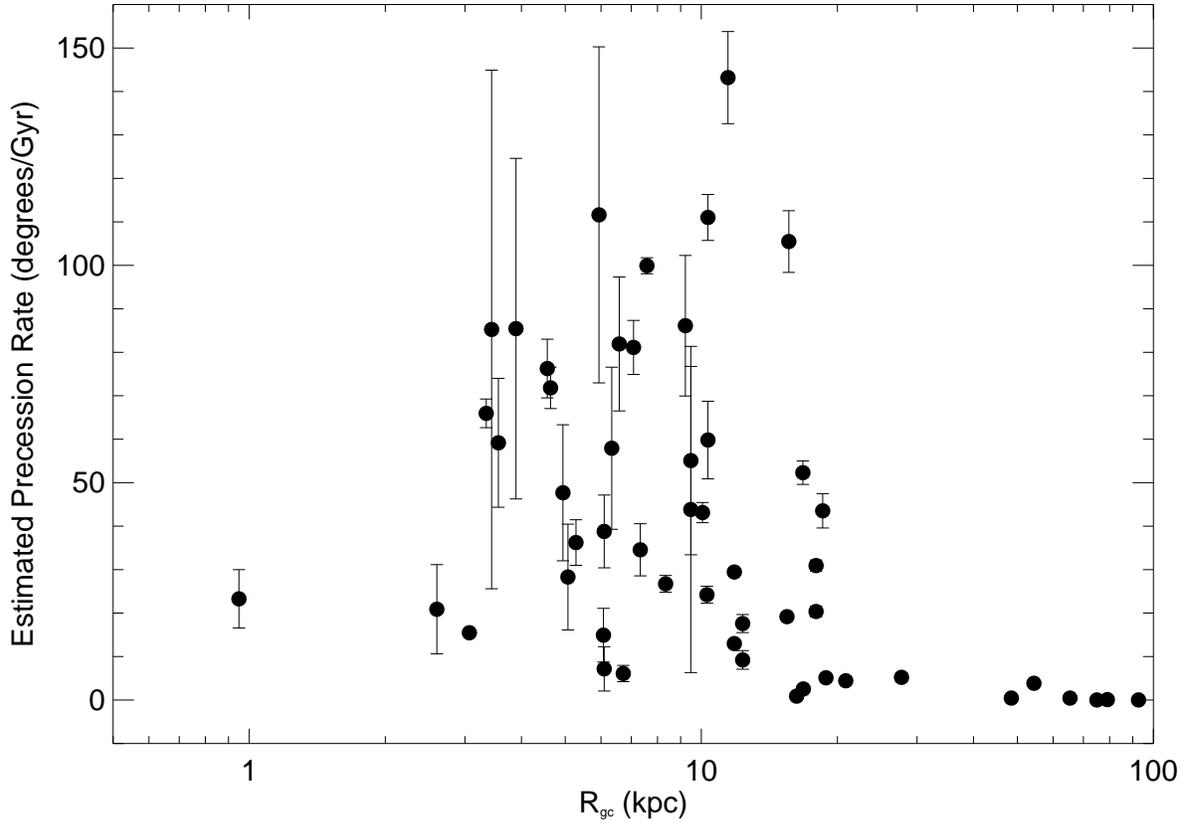}
\caption{An estimate of the rate of precession of the orbital poles of
our sample of clusters and DSGs with known proper
motions as a function of their Galactocentric distance.  Orbits
were integrated for 10 Gyr for each object in the
potential of \citet{kvj95}, and the pole was redetermined
each Gyr.  An angular separation was calculated between the
location of the pole in the current timestep and that from the previous epoch.
Plotted is the mean of these separations, with the error bars representing the
dispersion in the calculated values.   The data suggest it is a good assumption
that precession is not important for objects with $R_{gc} \gtrsim 20$ kpc, and
that there is a significant variation in the amount of precession for
objects at smaller $R_{gc}$.
\label{fig:precess}}
\end{figure}

In practice, however, we do not expect to find perfect coincidences
among the orbital poles of tidal remnant objects for a number of
reasons:

\begin{enumerate}

\item Even in a spherical potential the debris orbits will
be spread in energy about the disintegrating satellite's orbital
energy, with a typical scale \citep[see e.g.,][for a
discussion]{kvj98}
        \begin{equation}
                \Delta E= r_{\rm tide} {d\Phi \over dR} \approx
                \left( {m_{\rm sat} \over M_{\rm Gal}}\right)^{1/3} v_{\rm circ}^2
                \equiv f v_{\rm circ}^2
        \label{de}
        \end{equation}
where $r_{\rm tide}$ is the tidal radius of the satellite
\citep[see][]{king}, $\Phi$ is the parent Galaxy potential,
$v_{\rm circ}$ is the circular velocity of the Galactic halo, $m_{\rm sat}$ is
the satellite's mass, $M_{\rm Gal}$ is the mass of the parent
galaxy enclosed within the satellite's orbit, and
the last equality defines the {\it tidal scale} $f$.
This spread in energy translates to a characteristic
angular width $f$ to the debris.

\item If the parent Galaxy potential is not perfectly spherical, then
the satellite's orbit does not remain confined to a single plane.  For
example, differential precession of the orbital pole of debris from a
satellite may be induced, so remnant objects with slightly different
orbital energies and angular momenta that are found at different phases
along the parent satellite's orbit will have different orbital poles.
Moreover, \citet{helmi} point out that debris from inner halo objects
is not expected to retain planar coherence.  Indeed, some objects
currently found within $R_{gc} \sim 20$ kpc show chaotically changing
orbital planes \citep[e.g., the cluster NGC~1851; see][]{dd99b}.
Because the GCPF/ASPF technique depends on the orbital poles of debris
remaining approximately constant, differential precession and other
dynamical effects act to degrade the signal we seek.  As shown in
Figure \ref{fig:precess}, however, outside of $R_{gc} \sim 20$ kpc,
orbital pole drifting is a small effect, and not likely to affect
orbital pole alignments greatly.  Thus, the GCPF/ASPF technique should
be robust when probing the alignments of outer halo objects, however
for inner halo objects it will only be useful for relatively recent
disruptions.

\item Recent, high resolution N-body simulations of structure formation
suggest that the halo may be filled with dark matter lumps
\citep{kly99, moore} which could cause scattering of objects
away from the orbital plane. 

\item If the Galaxy's potential evolves significantly (e.g., through
accretion of objects or growing of the disk) this could cause initially
aligned orbital poles of objects to drift apart (see LB$^2$95 and
Zhao et al.\ 1999 for a more complete discussion).

\end{enumerate}

\begin{figure}
\plotone{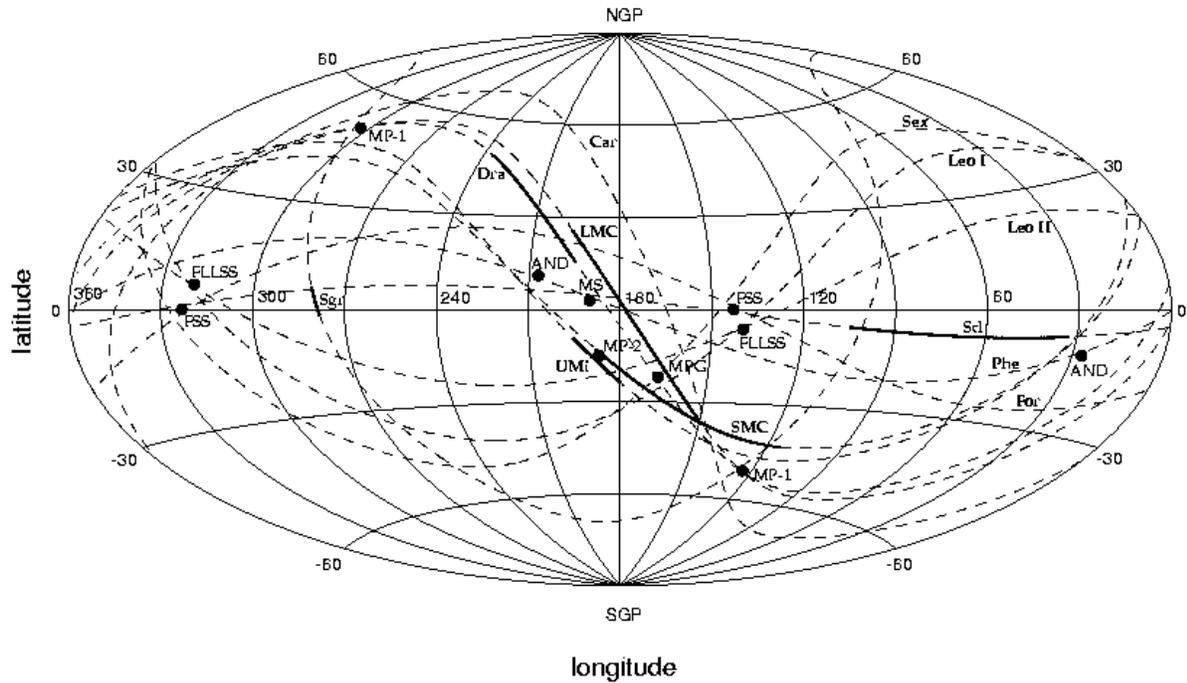}
\caption{This Aitoff projection in Galactocentric coordinates shows the
orbital pole families for the Galactic satellite galaxies \citep[see
also][]{srm96}.  The dashed
lines are great circle pole families (labelled in boldface with the
abbreviated galaxy name) that show all possible locations for an object's orbital
pole based solely on its Galactocentric radius vector.  Note the
multiple intersection point of the great circles of the ``Magellanic
stream Galaxies'' near $(l,b) = (140,-40)^{\circ}$.  The thicker lines
in this figure show the more restricted possible orbital pole locations
(``arc segment pole families'') for the six galaxies with published
space motions (plotted here are the ASPFs constructed with the
Schweitzer et al.\ (1997) proper motion for Ursa Minor, the Jones et
al.\ (1994) proper motion for the LMC, and the Irwin et al.\ (1996)
proper motion for the SMC, see Table \ref{pmtab}).  It is evident from the 
ASPFs that Draco and the LMC are likely on very similar
orbits, as are Ursa Minor and the SMC, and that all lie
remarkably near the GCPF intersection point.  Also, the positions of all of
the ASPFs show that nearly polar orbits are
preferred by Milky Way satellites. For reference, the filled circles
(labelled in plain text) indicate the positions of the poles of
previously proposed satellite alignments (see text).
\label{fig:dsphaitoff}}
\end{figure}

\noindent However, if the timescales for orbital pole drifting in the
outer halo are long enough, the true poles of tidal remnants should
remain relatively well aligned, and the GCPFs of these remnants may all
intersect within a small region on the sky (where ``small'' can be
estimated on the basis of the tidal scale, for example).  Indeed, the
argument may be turned around; given the various processes that tend to
cause drift among orbital poles, any well-defined, multiple GCPF
crossing point region that remains today is of particular interest.
For example, the GCPFs of the Magellanic stream DSGs (LMC, SMC,
Draco, and Ursa Minor) are nearly coincident, due to the spatial
alignment of these objects on the sky, and their GCPFs share nearly
common intersection points (dashed lines in Figure
\ref{fig:dsphaitoff}).  This nexus points to the oft-cited possibility
of a dynamical connection for these four objects.

However, while the GCPFs' nearly common crossing points indicate a {\em
possible} association, the true orbital poles of these objects can lie
anywhere on their respective Great Circle Pole Families and do not
necessarily need to lie near the crossing point.  Thus, nearly common
crossing point of the GCPFs of several DSGs does not indicate with any
certainty that the objects involved are truly the remnants of a past
tidal disruption, since the GCPFs contain no information about the
orbital energy or the magnitude and direction of the angular momentum
for each object.  Therefore, one may use common crossing points to
identify objects that are {\em potentially} dynamically linked, but
then other information, such as the orbital energy and momentum, must
be used eventually to check the likelihood of the dynamical
associations.  For example, in LB$^{2}$95, measurements of the radial
velocity of each satellite and an assumed model of the Galactic
potential were used to estimate the orbital energies and specific
angular momenta of their proposed candidate dynamical family members.
LB$^{2}$95 also assumed that the tidal elongations of the DSGs lie
along their orbital paths, which allowed them to hypothesize a true
pole location for each.\footnote[5]{In the particular case of the
Magellanic stream family, this assumption is supported, for example, by
the observation of \citet{oh} that the flattenings observed by Irwin \&
Hatzidimitriou (1993, 1995) in the outer parts of Draco, Carina, and
Ursa Minor appear to align with the plane defined by the Magellanic
Stream.}

Clearly, approximate knowledge of the {\em true} space motions of Milky
Way satellites is an improvement over the basic LB$^{2}$95 GCPF
technique, since one may constrain the true orbital pole to lie
somewhere along an ASPF.  Moreover, one can estimate the orbital energy
(within an assumed potential) and angular momentum directly, rather
than relying on assumed orbits.  For example, in Figure
\ref{fig:dsphaitoff}, we show the ASPFs (thick lines) for the Milky Way
DSGs, where the ASPFs are constructed using the space motion data
available in the literature \citep[see also][]{srm96}.  For reference
the poles of all previously proposed DSG/globular cluster alignment
planes, including Lynden-Bell's (1982) Magellanic stream (MS), Kunkel's
(1979) Magellanic Plane Group (MPG), and the MP-1 and MP-2 planes of
\citet{fusi}, are included.  The Fornax-Leo-Sculptor plane of
\citet{lb82} is illustrated with the ``FSS'' (LB$^2$95) point and the
``FLLSS'' point (Majewski 1994).  The Andromeda plane (AND) of
\citet{fusi} is also included for reference.  The space velocity data,
represented by the ASPFs in Figure \ref{fig:dsphaitoff}, support the
notion of a true ``dynamical group'' among the Magellanic stream DSGs
since their ASPFs do not appear to be randomly distributed around the
sky, but tend to lie remarkably near the multiple GCPF crossing point
region, which lies near the poles of the MS, MP-2, and MPG alignment
planes.  In addition, Figure \ref{fig:dsphaitoff} illustrates that
among DSGs with measured proper motions, all are on nearly polar orbits
(i.e.\ their orbital poles are near the Galactic equator).  Since polar
orbits are the least affected by precession, we expect that the ASPFs
for these objects have remained nearly constant over the lifetime of
each DSG.

Although knowledge of the space velocities of Galactic satellites
allows one to select objects with similar orbital dynamics with some
confidence, the current measurements of space velocities for most
satellites are not of the quality necessary to perform this task with
definitive results.  Since this limitation in data quality does not yet
allow the precise identification of dyamical families among the
Galactic satellite population, other information must presently be used
to reinforce the inclusion of (or to eliminate from consideration)
potential stream members identified by orbital pole alignments.
LB$^2$95 solved this dilemma by inverting the problem; objects were
selected with velocities and distances that produced radial energies
consistent with stream membership, and only then were the potential
orbital pole families of these objects searched for possible
alignment.

The goal of this work is similar to that of LB$^2$95, however, we
differ in not relying on the radial energy technique to discriminate
dynamically associated satellites from chance alignments, since 
this was thoroughly pursued in that paper.  Rather, we concentrate on the
more general view of what can be learned with full phase-space
information -- both with the current data and with an eye toward
refinements in the distances and space velocities of Galactic
satellites and clusters to be delivered by the astrometric satellite
missions \textit{SIM, GAIA, FAME,} and \textit{DIVA}. Therefore, our 
philosophy is to pursue a more liberal listing of possible associations
based on orbital polar alignments that can be tested with these future data.

\section{Case Study of the Sgr System}

Because Sagittarius is a paradigm for the type of tidally disrupted
system for which we are searching, it is worthwile to explore this example in
detail.  Others have proposed that Sgr is currently losing its globular
clusters to the Milky Way; for example, \citet{dCA95} argued that
Terzan~7, Terzan~8, and Arp~2 all belonged to Sgr and may be in the
process of being tidally removed from the galaxy \citep[see
also][]{ib95}, while \citet{dd00} proposed that Pal~12 was removed from
Sgr during a previous pericentric passage.  Since stellar debris from
Sgr has been identified at increasingly displaced positions on the sky
\citep{mmetal98, srm99, sdss, ibcarb} we investigate here the
possibility that there may be other captured Sgr clusters distributed
among the Galactic globular cluster population.

\begin{figure}
\plotone{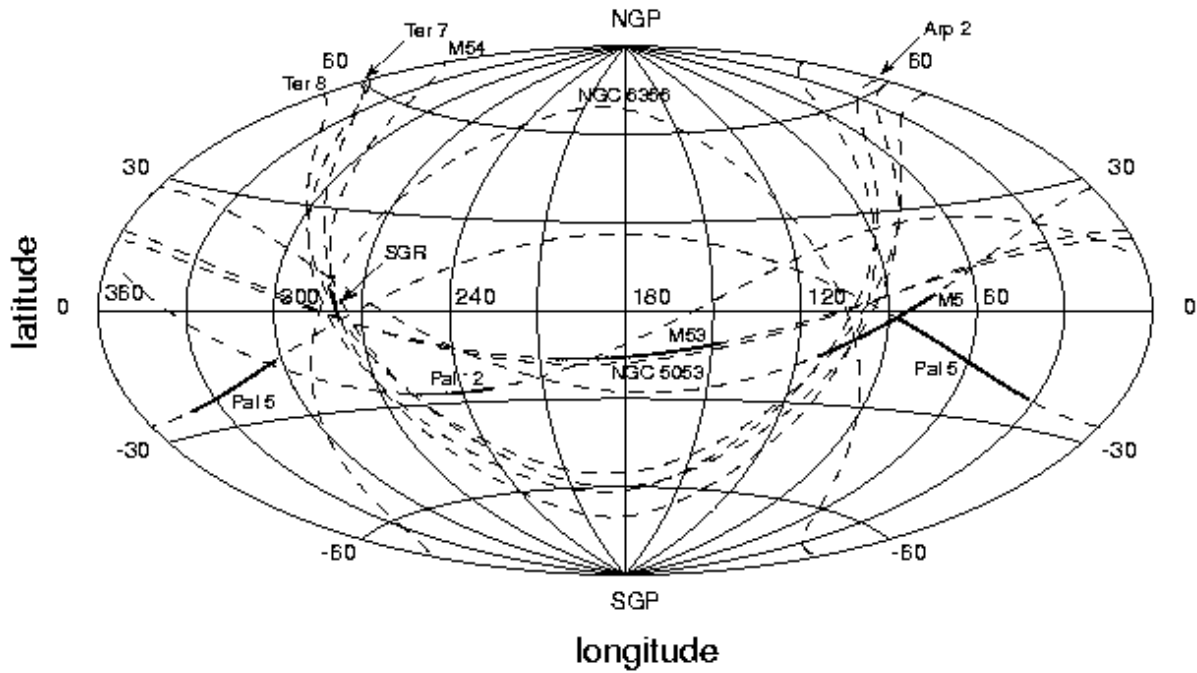}
\caption{An Aitoff projection in Galactocentric coordinates including
the pole families of the Sagittarius dwarf and those clusters
potentially associated with Sgr based on their GCPFs.  The dashed lines
are the GCPFs, while the thick, solid lines are the ASPFs for those
objects with measured proper motions (Pal~5 is shown with two ASPFs
derived from discrepant proper motions).  Note that all of the globular
clusters with proper motions appear to follow nearly polar orbits, as
do many of the DSGs (see Figure \ref{fig:dsphaitoff}), including
Sagittarius.  The orbital parameters of M53 and M5 are more similar to
those of Sgr than are those of Pal~5 (see Table \ref{sgrgctab}), however, the
physical properties of Pal~5 (metallicity, luminosity, concentration)
are very similar to the other Sgr clusters.  \citet{dd00} propose that 
Pal~12 is an Sgr cluster that was stripped on a previous pericentric passage. 
Although its pole does not align with that of Sgr, orbital integrations
by \citet{dd00} show that Pal~12 and Sgr were more closely aligned in phase space in
the past.
\label{fig:sgr}}
\end{figure}

In Figure \ref{fig:sgr}, we show the Sgr ASPF along with the GCPFs for
a sample of globular clusters selected as potential Sgr debris based on
the proximity of their GCPFs to the Sgr ASPF.  The globular clusters in
the sample, M53, NGC~5053, Pal~5, M5, NGC~6356, M54, Terzan~7,
Terzan~8, and Arp~2 were chosen to satisfy two criteria:  (1) their
GCPFs come within $5^{\circ}$ of the Sgr ASPF (see \S 3.2 for a
discussion of the expected angular width of tidal debris derived from
$\Delta E$, where $m_{Sgr}$ was assumed to be $< 10^8 M_{\sun}$), and
(2) $6.0 \leq R_{gc} \leq 36.0$ kpc.  Of this group, NGC~6356
($[\textrm{Fe/H}] = -0.5$) has properties similar to the \citet{burk97}
metal-enriched, inner halo population, which they argue formed during
the collapse phase in the Galaxy's formation.  However, an alternative
explanation for NGC~6356's halo-like orbit could be accretion from a
Galactic DSG; we note that its metallicity is similar to that of the
Sgr cluster Terzan~7.  On the other hand, M53 and NGC~5053 are more
metal poor than the previously identified Sgr clusters
($[\textrm{Fe/H}] = -1.99$ and $-2.29$, respectively).  Of the
remaining candidate Sgr clusters, four are those previously proposed to
be Sgr clusters (M54, Ter~7, Ter~8, and Arp~2); the other two are Pal~5
and M5, which are both second parameter, red horizontal branch
clusters.  The physical and orbital parameters (orbital energy, orbital
angular momentum, $R_{gc}$, $R_{apo}$, $R_{peri}$, $[\textrm{Fe/H}]$, $M_V$, and
concentration parameter) are tabulated in Table \ref{sgrgctab} for the
known Sgr globular clusters (bottom of table) and for the new candidate
Sgr clusters presented here (top of table).

The ASPFs of the three clusters in Figure \ref{fig:sgr} with measured
proper motions (M53, Pal~5, and M5) are plotted as thick, solid lines.
Pal~5 is shown with two ASPFs, since the \citet{kmc01p5} and
\citet{scholz98} proper motions are so discrepant that the resultant
ASPFs lie almost $180^{\circ}$ apart.  It is interesting to note that
like Sgr, M53, Pal~5, and M5 are apparently on nearly polar orbits.
Perhaps these clusters were Sgr clusters, but precession has caused
their poles to drift from that of Sgr?  
These three clusters are at
$R_{gc} < 20$ kpc, where precession effects are more significant
(Figure \ref{fig:precess}
shows drifts of up to 50$^{\circ}$ at 16 kpc), however
nearly polar orbits generally precess more slowly than orbits with
smaller inclinations.  The orbital parameters suggest that Pal~5 is on
an orbit unlike that of Sgr.  With either proper motion, the orbital
energy and angular momentum of Pal~5 differ from Sgr, but there is
enough uncertainty in the differences that an Sgr debris orbit can not
be completely ruled out.  However, the orbital energy and angular
momentum of M53 are very similar to Sgr ($L = 47\pm23$ and $44\pm5$
$10^{2}$ kpc km/sec respectively and $E_{orb} = -4.4\pm2.5$ and
$-4.4\pm0.6$ $10^{4}$ km$^2$/sec$^2$ respectively).  For M5 the values
are more discrepant ($L = 14\pm5$ and $E_{orb} = -1.8\pm2.3$), however,
the proper motion of M5 is of lower precision than either that of Pal~5
or M53.

The available data do not allow us to definitively identify any of
these clusters as captured Sgr clusters, however several of the
candidates are similar enough to the system of Sgr globular clusters to
warrant further investigation.  Dinescu et al.\ (2000) argue for Pal~12
as an Sgr cluster due to its dynamics, and also because its
metallicity, mass, and concentration are similar to the other Sgr
clusters.  Pal~5 has a metallicity, mass, and concentration similar to
Pal~12, Ter~7, Ter~8, and Arp~2, however its orbit seems too different
from that of Sgr to be definitively considered a captured Sgr cluster.
M53 is more metal poor, more massive, and more centrally concentrated
than all the Sgr clusters except for M54, which is postulated to be
the nucleus of Sgr, however M53's dynamics seem to match well
with Sgr.  Finally, M5 has a metallicity similar to the Sgr clusters,
dynamics that may be consistent with Sgr, yet it too is more massive
and centrally concentrated than Pal~12, Ter~7, Ter~8, and Arp~2.  
This case study of Sagittarius demonstrates the potential
of the pole analysis 
technique to uncover dynamical associations and presents several new candidates
(Pal~5, M53, M5, and NGC~6356) for an extended Sgr
dynamical family.

\section{Great Circle Pole Family Analysis}

Now, we search the entire sample of halo globular clusters and 
DSGs for nexuses of multiple GCPFs similar to that seen
among the Magellanic stream DSGs (Figure \ref{fig:dsphaitoff}).
Although the core of the GCPF analysis presented here is not
significantly different than that of LB$^2$95, our study differs in
that we (1) investigate the possible dynamical association of various
distinct subpopulations of globular clusters with Milky Way 
DSGs and (2) compare the probability of all potential
dynamical associations in a statistical, rather than anecdotal,
manner.

We analyze the GCPFs of a sample of objects in the following way:

\begin{enumerate}

\item For each pair of objects in the sample, calculate the two points
along their respective GCPFs where there is an intersection.  We remind
the reader that this introduces a redundancy due to symmetry around the
antipodes, but this redundancy has been taken into account during the
analysis.

\item Calculate the angular distance (along the connecting great
circle, i.e., the minimum angular distance) between each crossing point
and every other crossing point.

\item To assess the true statistical significance of clustering among
the GCPF crossing points, we calculate the two point angular
correlation function $w(\theta)$ for the crossing points of various
subsamples and compare to the results for other subsamples.

\end{enumerate}

There are various ways to estimate $w(\theta)$, and we adopted the
technique for calculating the $w_{1}(\theta)$ estimator outlined in
\citet{lsz93}.  The $w_{1}(\theta)$ estimator compares the distribution
of angular separations of data/data (DD) pairs to that of random/random
(RR) pairs (a discussion of the random data generation is included in
Appendix A).  One cannot use other estimators (such as the
$w_{4}(\theta)$ estimator; Landy \& Szalay 1993) that rely on, for
example, the comparison of DD pairs to a cross-correlation of the data
points to random points (DR pairs) due to the nature of GCPF crossing
points.  The problem lies in the fact that there is an intrinsic
correlation in crossing point data because all points on the celestial
sphere are not equally likely to have a crossing point: Only those
points that lie along two great circles may be crossing points.  The
same intrinsic correlation of crossing point location applies to any
randomly generated set of GCPF crossing point data, as long as the
crossing point distribution is derived {\em after generating a random
set of constraining great circles}, rather than simply generating random
{\em crossing points} that lie {\em anywhere} on the celestial sphere.  For
example, the
orientations of the great circles in the real dataset are completely
uncorrelated to the positions of the great circles in the random
dataset, so DR cross-correlations do not have the intrinsic correlation
found in the DD and RR data and false signal amplitudes will be
generated in comparison of DD to DR pairs.  In such a misapplication
of the technique, then, since the positions of the
great circles in the real data are independent of the positions in the
randomized data, the DD/DR estimators would measure the amplitude of the
clustering as well as the amplitude of the correlation in the positions
of great circles in the real data, and therefore artificially inflate the
amplitude of $w(\theta)$.  On the other hand, the appropriate, DD/RR, 
estimators measure
only the amplitude of the clustering in the data, since the same
intrinsic correlation is contained in both the DD and RR pairs
(if the great circles are randomized fairly; see Appendix A) and
falls out when the ratio is taken.

The total sample of globular clusters and Milky Way DSGs 
is very large.  Since there will be $(N_{objects})(N_{objects}-1)/2$
crossing points, the signature of a true dynamical grouping can be lost
in the ``noise'' of random crossing points.  Moreover, we suspect that
there may be some correlation of physical characteristics in the debris
from a common progenitor.  Thus, variously selected samples of similar
objects may show a higher degree of orbital coherence.  For example,
the \citet{SZ} fragment accretion hypothesis was motivated by the
predominance of the second parameter effect in the outer halo globular
clusters.  We therefore can hope to improve the signal-to-noise in
potential dynamical families with judicious parsing of the sample into
physically interesting subsamples.

It has been argued that selecting globular clusters by metallicity
\citep[e.g.,][]{rp84}, by horizontal branch morphology (e.g., Zinn 1993,
1996), or by Oosterhoff class \citep{vdbgh} will separate the
globular cluster population into distinct sub-populations with
different kinematical properties that may in some cases be indicative
of an accretion origin.  We have attempted to reproduce some of these
divisions to determine if one particular sub-population has a greater
incidence of GCPF crossing point clumping than the others.  The results
are summarized in the following sections.

\subsection{Zinn RHB vs. Zinn BHB/MP Globular Clusters} 

After dividing globular clusters into two types using the [Fe/H] vs. HB
morphology parameter diagram \citep{lee94}, \citet{zinn93a} found that
kinematic differences exist between the ``Old Halo'' (BHB/MP) and
``Younger Halo'' (RHB) populations, evidence that supports the
accretion model of the second parameter, RHB clusters, which
predominate in the outer halo.  \citet{srm94} has shown evidence for a
spatial association between a sample of Zinn RHB globulars
(specifically, those with the reddest HBs) and the
Fornax--Leo--Sculptor stream of DSGs, which
suggests that these may be related to a common Galactic tidal
disruption event.  We support these spatial and kinematical
associations among the RHB globulars with the results of our GCPF
technique.  Using the metallicity and HB morphology data from the
\citet{mwgc} compilation and adopting the same partitioning scheme as
in the \citet{zinn93a} paper, we separated all non-disk Milky Way globular
clusters into the RHB and BHB/MP types.  In so doing, we take into account
the recommendations of \citet{dCA95} to include Ter~7, Pal~12, and Arp~2
in the RHB group.  Since we are interested mainly in the outer halo,
where dynamical families are expected to be best preserved from phase
mixing \citep{helmi}, and where there is minimal contamination by disk
clusters, we also impose an $R_{gc} > 8$ kpc cutoff for both groups.
The final samples, listed in Table \ref{hbtab}, contain 22 BHB/MP
globular clusters and 26 RHB globular clusters.  We analyze both
samples twice: either including or excluding the Milky Way 
DSGs.

\begin{figure} 
\plotone{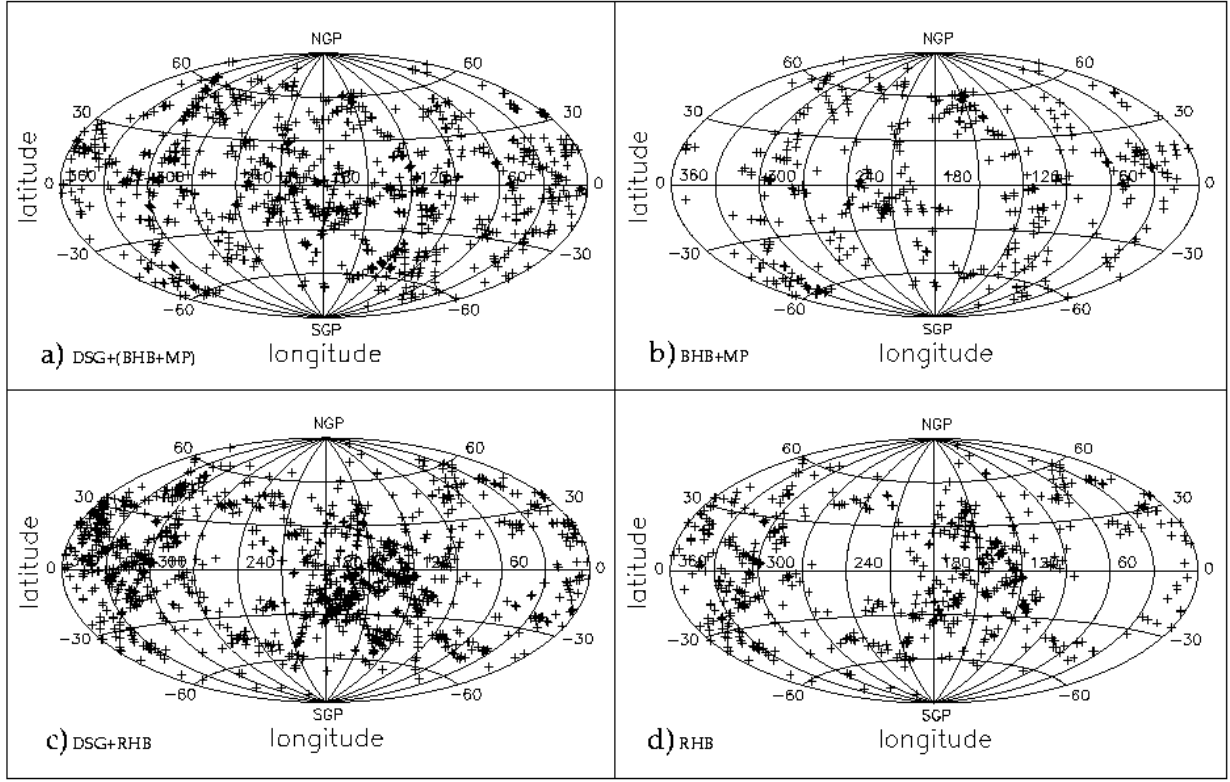}
\caption{Galactocentric distributions of crossing points
of pairs of great circle pole families.  Panel (a) shows the crossing
points for the sample that includes all Milky Way 
DSGs and the Zinn BHB/MP globular clusters with $R_{gc} > 8$
kpc.  Panel (b) is the distribution for the same globular cluster
sample as in panel (a), but with the DSGs removed.  Panel (c)
shows the distribution of crossing points for the sample of Zinn RHB
globular clusters with $R_{gc} > 8$ kpc and the Milky Way 
DSGs.  Panel (d) is the distribution for the same globular
cluster sample as in panel (c), but with the DSGs removed.
Note the large cluster of crossing points in panel (c) that is located
near the location of intersection of the pole families of the
Magellanic stream and FL$^{2}$S$^{2}$ stream DSGs shown in Figure
\ref{fig:dsphaitoff} (the distribution of points is symmetric about
180$^{\circ}$ since each pair of great circles has two intersection
points; so the clump near $(l,b)=(345,25)^{\circ}$ is an anitpodal reflection of
the clump we refer to here).  It is especially interesting to note that
there is still a large cluster of crossing points in this same general
area in the sample that includes only the Zinn RHB globulars (panel
(d)), while this area is empty in the Zinn BHB/MP globular
cluster sample (panel (b)), and, moreover, there are no clusters of
BHB/MP crossing points of the magnitude of that shown by the RHB
globular clusters. 
\label{fig:crosspts}}
\end{figure}

From a comparison of the distribution of crossing points for the Zinn
RHB globular clusters$+$DSGs to the crossing points for the Zinn BHB/MP
globular clusters$+$DSGs (Figure \ref{fig:crosspts}), it is clear
that there are several large groupings in the RHB$+$DSG sample that are
absent or are much smaller in the BHB/MP$+$DSG sample.  It is
especially interesting to note that the largest grouping found in both
datasets is located where the GCPFs of the FL$^{2}$S$^{2}$ DSGs cross
those of the Magellanic stream Group DSGs, near $(l,b) =
(165,-25)^{\circ}$ (see Figure \ref{fig:dsphaitoff}).  However, when one
looks at the crossing points of only the RHB or BHB/MP globular
clusters without including the GCPFs of the DSGs, there is still a
large grouping of crossing points in the RHB globular cluster data, but
{\em no} crossing points in this region in the BHB/MP globular
cluster data.  This indicates that this particular excess of crossing
points near $(l,b) = (165,-25)^{\circ}$ in the BHB/MP$+$DSG sample is
entirely due to the DSGs, while in the RHB$+$DSG sample there are a
large number of globular cluster/globular cluster crossing points also
found in the $(165,-25)^{\circ}$ region.  Moreover, there appears to be
no clustering of GCPF crossing points in the BHB/MP globular cluster
sample that is anywhere near the size or density of that seen in the
RHB cluster sample.

\begin{figure}
\epsscale{0.6}
\plotone{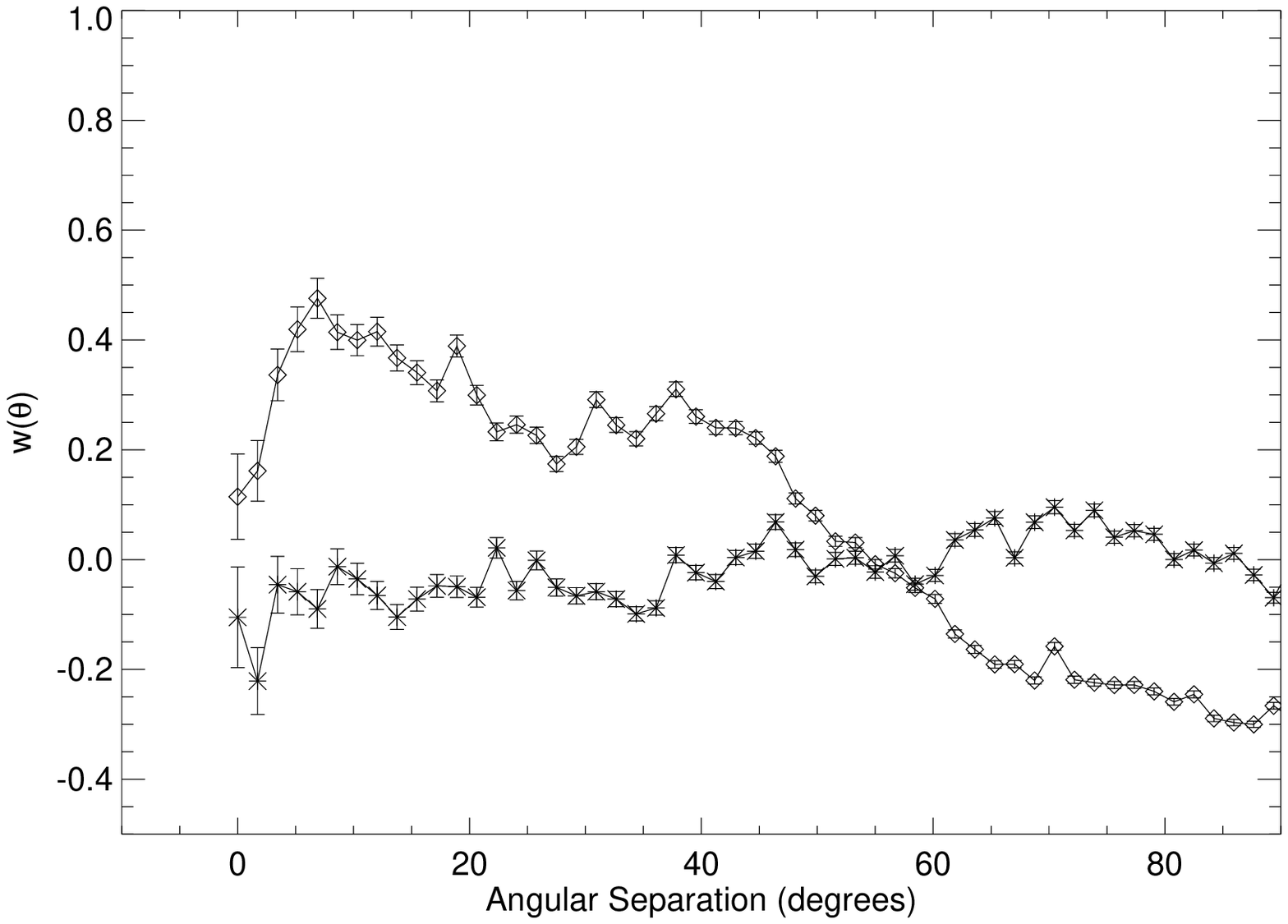}
\plotone{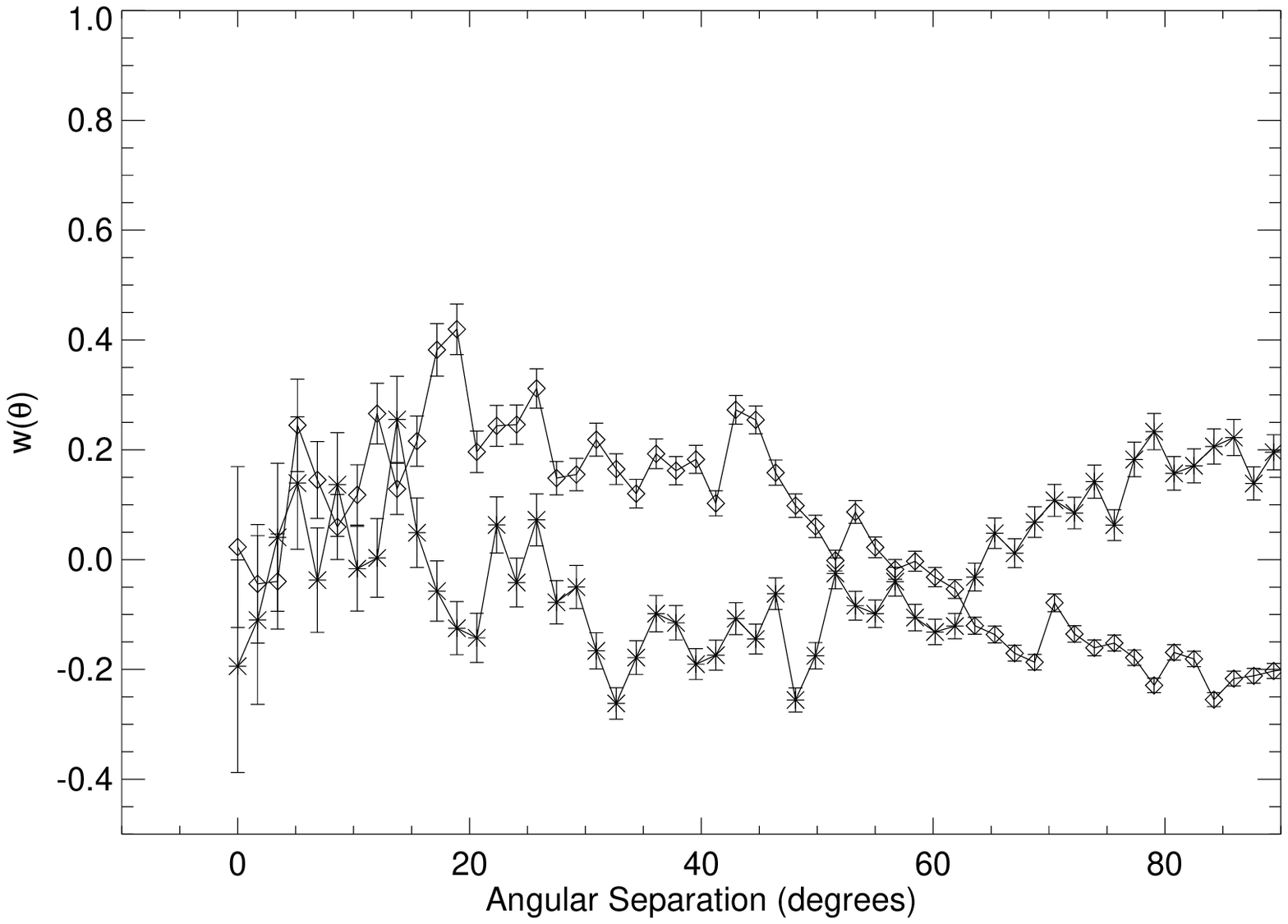}
\caption{Upper panel:  Angular two-point correlation function analysis of the
distribution of crossing points in the Zinn RHB globular
cluster$+$DSG (panel c Figure \ref{fig:crosspts}) sample versus the
Zinn BHB/MP globular cluster$+$DSG (panel a Figure
\ref{fig:crosspts}) sample.  The points marked with open diamonds
correspond to the Zinn RHB globular cluster$+$DSG sample, and the
points marked with asterisks correspond to the Zinn BHB/MP globular
cluster$+$DSG sample.  Note that for all scales $\lesssim
55^{\circ}$, the Zinn RHB globular clusters$+$DSG sample shows a
significantly greater amplitude than does the Zinn BHB/MP globular
cluster$+$DSG sample.  Lower panel:  Angular two-point correlation function
analysis of the distribution of crossing points for the Zinn RHB globular
clusters alone (panel d Figure \ref{fig:crosspts}) versus the Zinn BHB/MP
globular clusters alone (panel b Figure \ref{fig:crosspts}).  Although the 
signal to noise is poorer in this figure than in the upper panel where
the DSGs have been included, there does
appear to be some evidence that the RHB globular cluster crossing points
have a larger clustering amplitude than do those of the BHB globular clusters.
\label{fig:2pt}}
\end{figure}

The two point angular correlation function calculation shows
that the RHB$+$DSG sample has a larger $w(\theta)$ amplitude
than that of the BHB/MP$+$DSG sample out to scales of 
$55^{\circ}$ (Figure \ref{fig:2pt}, upper panel).  The $w(\theta)$ amplitude for the 
BHB/MP$+$DSG sample is actually consistent with it being a random
distribution (i.e. DD/RR$-1 \sim 0$).  This is statistical
verification of what one sees by eye: The crossing points for the
RHB globular clusters $+$ DSGs are more clumped than
those of the BHB/MP globular clusters$+$DSGs.  If one
looks at the $w(\theta)$ amplitude for the globular cluster samples
alone (Figure \ref{fig:2pt}, lower panel), 
the sample size is sufficiently small that there is a large
overlap in the error bars for points $< 15^{\circ}$, 
so it is difficult to prove definitively that the
RHB globular clusters taken alone are more clumped than the 
BHB/MP globular clusters.  However, there does appear to be a larger
$w(\theta)$ amplitude for the RHB globular clusters than for the
BHB/MP clusters, particularly for $\theta \sim 15^{\circ} - 20^{\circ}$.

\begin{figure}
\epsscale{1.0}
\plotone{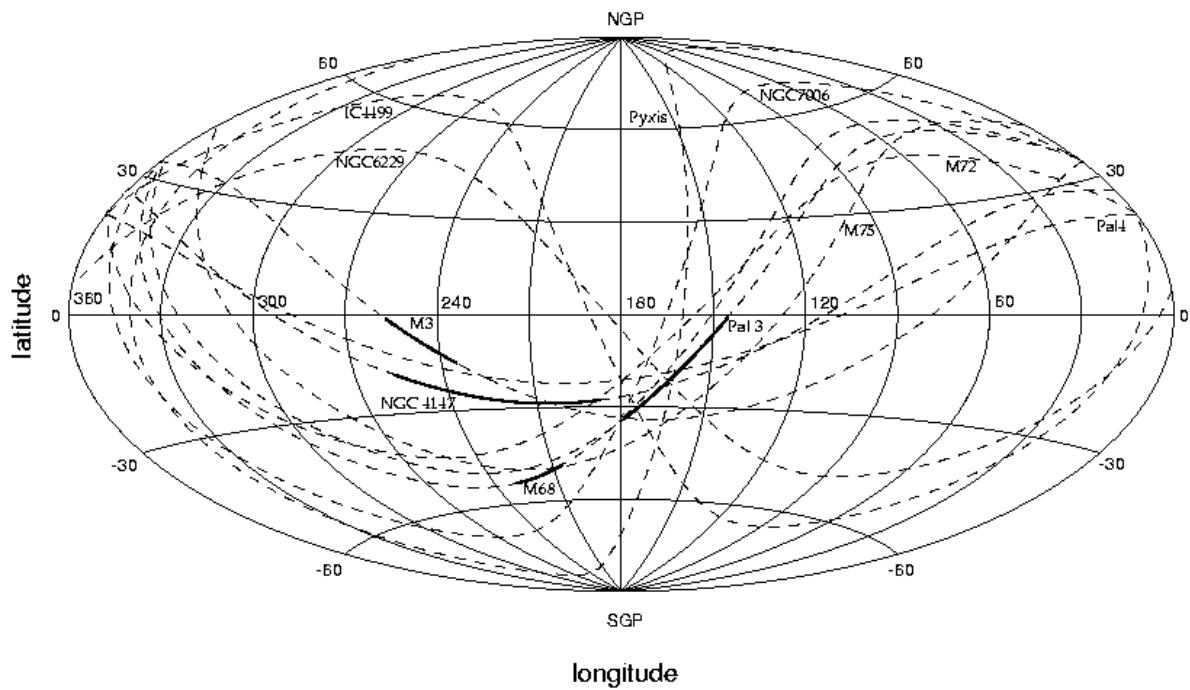}
\caption{The pole families of the 11 globular clusters that produce the
excess clustering (Figure \ref{fig:2pt}) in the crossing point
distribution for RHB (second parameter) $R_{gc} > 8$ kpc globular
clusters.  The dashed lines are the great circle pole families, the
solid lines indicate the better constrained, arc segment pole families
for those four globular clusters in this group with measured proper
motions.  Comparing this to Figure \ref{fig:dsphaitoff}, it is clear
that the multiple intersections of the pole families of these objects
are in the same part of the sky where the pole families of the
Magellanic stream galaxies cross with those of the FL$^{2}$S$^{2}$
galaxies.  One expects that at least some of these 11 may in fact be
associated with one of these two groups of DSGs.  The arc
segment pole families for the four globular clusters with measured
proper motions show that only one of these four has its true orbital
pole near enough to the ASPFs of the Magellanic stream group ASPFs to
have been agglomerated into the Magellanic stream Group by the cluster
analysis algorithm (see \S6.2).   However, the globular cluster ASPFs
are surprisingly close to the ASPFs of the SMC and Ursa Minor, and
since the proper motions of M3 and NGC~4147 have large
errors associated with them, we do not rule out this tentative association
completely.  Of the remaining seven globular clusters without
proper motions, four of these are in the outermost halo ($R_{gc} > 25$
kpc, Pyxis, NGC~6229, NGC~7006, and Pal~4), and are therefore
candidates to have been accreted into the Milky Way halo.
\label{fig:pyxisetc}}
\end{figure}

The majority of the $w(\theta)$ clustering signal comes from the large
clump of crossing points near $(l,b) = (165,-25)^{\circ}$.  This group
of crossing points near the equator supports the notion that polar
orbits are preferred not only by the DSGs (a notion that is verified by
the actual orbital data we have for some of the DSGs), but perhaps by
the second parameter clusters (which we find potentially to be
associated with these DSGs) as well.  Applying the cluster finding
algorithm described in \S6.2 to the distribution of crossing points, we
find that the GCPFs of the following globular clusters create the
excess of crossing points near the equator:  Pyxis, IC~4499, Pal~3,
Pal~4, M3, M68, M72, M75, NGC~4147, NGC~6229, and NGC~7006.   If we
apply a less conservative angular cutoff when we determine which
crossing points contribute to this excess, we find that Pal~12, AM1,
NGC~2808, Rup~106, and NGC~6934 also contribute to the size of the
clump of crossing points.  Figure \ref{fig:pyxisetc} shows a plot of
the pole families for the 11 globular clusters selected with the
conservative partition; a discussion of the ASPF distribution of the
four globular clusters in this group with measured proper motions is in
\S6.3.  Although the DSGs have been previously separated into two
groups, the Magellanic stream group and the FL$^2$S$^2$ group, we must
note here that the GCPFs of the Magellanic stream galaxies intersect
those of the FL$^2$S$^2$ galaxies in the region near $(l,b) =
(165,-25)^{\circ}$.  Thus, the large clump of crossing points (which
appears to be made of two subclumps; one due to the Magellanic stream
group galaxies and one due to the FL$^2$S$^2$ galaxies) contains
crossing points derived from the GCPFs of all of the globular clusters
listed above and both Magellanic stream and FL$^2$S$^2$ galaxies.
Therefore it is possible that many of these globular clusters belong to
either the Magellanic stream or the FL$^2$S$^2$ groups.  Of this group
of 11 globular clusters, Pal~3, Pal~4, and M75 are part of the group of
proposed FL$^2$S$^2$ globular clusters of \citet{srm94}, IC~4499 is
part of the MP-1 plane of \citet{fusi}, and \citet{wekdem76} include
Pal~4 and NGC~7006 in their Magellanic Plane group.

\citet{fw80} found that the radial velocities of 66 globular
clusters with $R_{gc} < 33$ kpc are consistent with a systemic rotation
around the Galactic pole of 60$\pm$26 km/sec.  More recently, Zinn
(1985, 1993) used the \citet{fw80} technique to show that the
globular clusters with [Fe/H] $> -0.8$ have disklike rotation
velocities, while the more metal-poor globular clusters have a
marginally significant net rotation with a large velocity dispersion.
In all of these studies, the globular clusters were assumed to have a
systemic rotation around the same axis as the Milky Way's disk, the
magnitude of which could be estimated from the component of the
rotation that lied along the line of sight to each cluster.  However,
our crossing point analysis suggests that a large group of globular
clusters may be following {\em polar} orbits, similar to the
trend of Milky Way satellite galaxies.   A measurement made with the \citet{fw80}
technique of a statistically significant systemic rotation
along this nearly polar orbital path for the sample of Zinn RHB
globular clusters we list above would strengthen our case for labelling
this sample as a potential dynamical group.  Unfortunately, such an
analysis yields little leverage on the problem.  In the case of orbits
flattened near the Galactic plane, which have axes of rotation near the
Galactic $Z$ axis, the Sun's position 8.5 kpc from the Galactic center
along the Galactic $X$ axis is fortuitous, since for these objects some
component of their systemic rotational velocity around the Galactic $Z$
axis is along our line of sight.  But, polar orbits that are oriented
with a rotation axis nearly aligned with the Galactic $X$ axis can not
be analyzed very well with the Frenk \& White technique, because there
is little or no parallax between the solar position with respect to the
Galactic center and the axis of rotation.  An attempt was made to
measure a systemic rotation around $(l,b)=(165,-25)^{\circ}$ for the
objects listed above, however due to small number statistics and the
small angle to the line of sight, a statistically insignificant result
was obtained.

\subsection{Metallicity selected subsamples} 

Using the \citet{fw80} technique, Rodgers \& Paltoglou (1984, hereafter
RP84) found that the sample of Galactic globular clusters with
metallicities\footnote[6]{RP84 used globular cluster metallicities from
three sources, \citet{zinn80}, \citet{hr79}, and \citet{kraft79}.  They
converted the data from all three sources onto a single system defined
by the \citet{zinn80} metallicities.  However, since they did not
publish the details of their calibration, their metallicity scale, and
in turn the specific clusters in each subsample are unknown.} in the
``window'' from $-1.3 > [\textrm{Fe/H}] > -1.7$ display a net
retrograde rotation,  while both higher and lower metallicity samples
showed net prograde rotations.  From this RP84 concluded that the
clusters in the intermediate metallicity sample may have derived from
an accretion event.  If so, we might expect to see corresponding
signals in our crossing point analysis.  We therefore separated the
$R_{gc} > 8$ kpc sample of Galactic globular clusters into three
metallicity selected subsamples:  $[\textrm{Fe/H}] > -1.3$, $-1.3 \geq
[\textrm{Fe/H}] \geq -1.7$, and $[\textrm{Fe/H}] < -1.7$ using abundances from
\citet{mwgc}.  The analysis here differs somewhat from that of RP84
because:  (1) our $R_{gc} > 8$ kpc restriction reduces the high
metallicity subsample to only 7 objects, so the intermediate
metallicity sample is only compared to the low metallicity sample, and
(2) modern metallicity values were used to divide the globular clusters
into subsamples, so the samples presented here are likely to be
different than those used in the original study.

The results of our analysis show that there is no significant excess
clustering in one sample relative to the other.  This is seen by
inspection of the crossing points, as well as in the amplitude,
$w(\theta)$, of the two point angular correlation function.  
Therefore, one can conclude that there is no correlation between 
metallicity (specifically in these particular metallicity bins) 
and similarity in orbit to the Milky Way DSGs.

We have examined the rotation sense of the orbits of the globular
clusters that fall in the $-1.3 > [\textrm{Fe/H}] > -1.7$ metallicity
window and also have published proper motions.  Of these 15 globular
clusters, only five are following retrograde orbits.  For our low
metallicity sample, eight of the 21 have published proper motions.  Of
these eight, three are on retrograde orbits.  So in both the low and
intermediate metallicity samples, approximately the same percentage of
globular clusters are following retrograde orbits.  With the
\citet{fw80} statistical technique, RP84 identified seven globular
clusters in their $-1.3 > [\textrm{Fe/H}] > -1.7$ sample as having the
largest retrograde motions, however only one of the three of these with
proper motion determinations is actually measured to be following a
retrograde orbit (NGC~6934).   Dinescu et al.\ (1999b) points out that
these three clusters in the RP84 ``retrograde'' sample with measured
space motions span a large range in orbital angular momentum and have
very different orbits, indicating a very low probability that they are
daughters of a single parent object.

\subsection{Galactocentric radius slices}   

Since the DSGs of the Milky Way (except Sagittarius) all presently lie
at $R_{gc} > 25$ kpc, it is natural to assume that the globular
clusters farthest from the Galactic center may have the highest
probability of having originated in tidal interactions between the
Milky Way and one of the DSGs.  For this reason, the sample of globular
clusters was divided into two samples, one with $(8 < R_{gc} < 25)$ kpc
and the other with $R_{gc} > 25$ kpc.   All of the Galactic satellite
galaxies were included with both samples of globular clusters.

\begin{figure}
\plotone{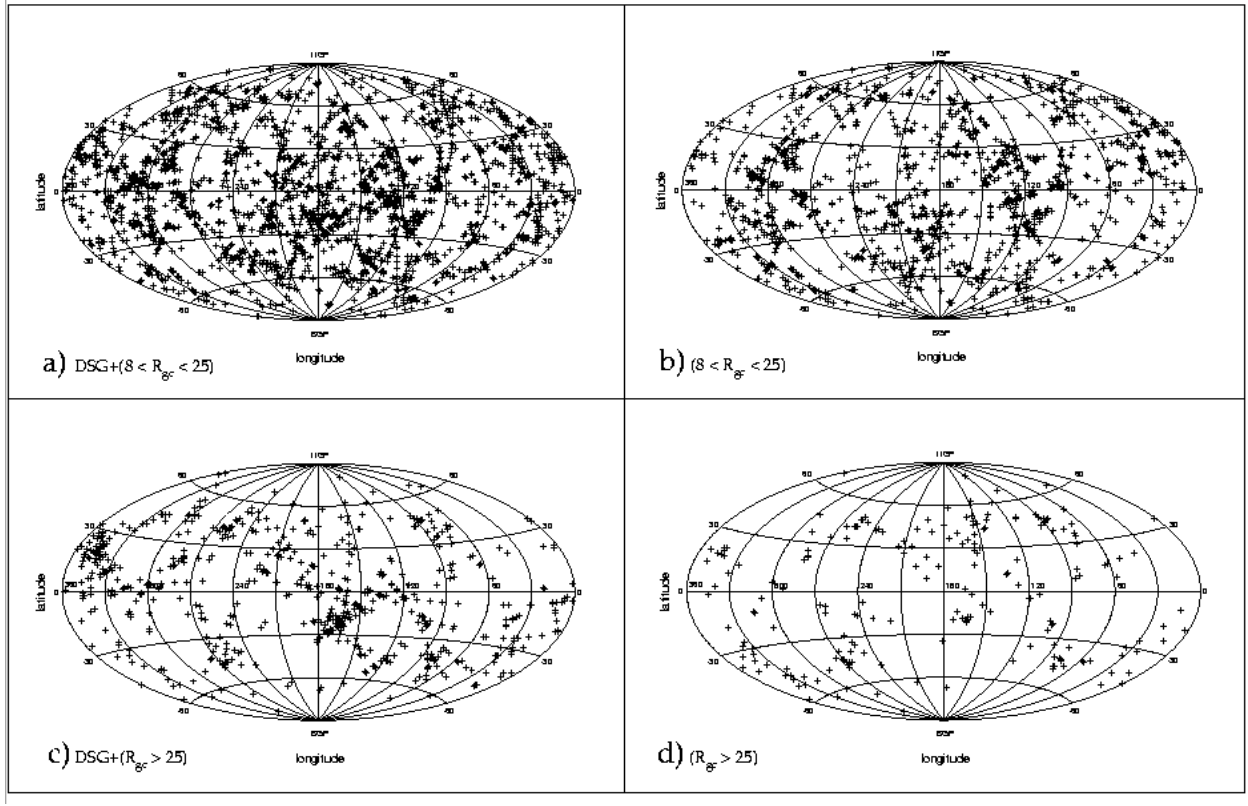}
\caption{GCPF crossing point distributions for the $(8 < R_{gc} < 25)$
kpc globular cluster$+$DSG sample (panel a) and the $(R_{gc} > 25)$ kpc globular
cluster$+$DSG sample
(panel b).  The distribution is mostly isotropic in panel (a),
however there is a clump near $(l,b) = (165,-25)^{\circ}$ due to the
intersection of the DSGs' GCPFs.  In the distant outer halo
sample, there is clearly no other clump of GCPF crossing points with a
similar size as the one near $(l,b) = (165,-25)^{\circ}$.  This clump
is due to the GCPF intersections of the Milky Way 
DSGs and the $R_{gc} > 25$ kpc globular clusters Pal~4, Pyxis,
NGC~6229, and NGC~7006.
\label{fig:2pangcpf}}
\end{figure}

If one considers the ($8 < R_{gc} < 25$ kpc globular cluster)$+$DSG
sample on its own, the number of GCPF crossing points is high since the
sample is large.  However, the distribution of the crossing points
(Figure \ref{fig:2pangcpf}) for this sample is fairly isotropic.  There
is some excess of GCPF crossing points near the Magellanic stream
intersection point, but as in the case of the BHB/MP globular
clusters discussed in \S5.1, this excess is entirely due to the DSGs.
The distribution of GCPF crossing points for the $R_{gc} > 25$ kpc
globular cluster$+$DSG sample (Figure \ref{fig:2pangcpf}) is more
sparse, however we again find the densest group of points to be in the
same region as when we considered all Zinn RHB globular
clusters$+$DSGs, near $(l,b)=(165,-25)^{\circ}$.  The difference in
clustering amplitude between these two samples is verified
statistically; comparing the $w(\theta)$ amplitude for the two samples
(Figure \ref{fig:2pt2}), we find that for the ($8 < R_{gc} < 25$ kpc)
globular cluster$+$DSG sample, $w(\theta) \sim 0$ over the same range
in $\theta$ represented in the upper panel of Figure \ref{fig:2pt}
($0^{\circ} < \theta < 50^{\circ}$), while for the $R_{gc} > 25$ kpc
globular cluster$+$DSG sample $w(\theta)$ is approximately equal to
that for the Zinn RHB globular cluster$+$DSG sample over the range
$0^{\circ} < \theta \lesssim 15^{\circ}$.  This result suggests that
the majority of the clumping among the GCPF crossing points at angular
scales expected for tidal debris ($\lesssim15^{\circ}$) is due
primarily to the DSGs and the $R_{gc} > 25$ kpc globular clusters
(which are dominated by RHB clusters).  The $w(\theta)$ amplitude is
large only on scales $\lesssim15^{\circ}$ because the $R_{gc} > 25$ kpc
limit seems to exclude the the FL$^{2}$S$^{2}$ subclump that can be
seen in Figure \ref{fig:crosspts}, with the result that the scale over
which the correlation amplitude remains significant is reduced.  This
would seem to indicate that the more distant RHB clusters are more
likely to be Magellanic stream members, while the $R_{gc} < 25$ kpc RHB
clusters associate with comparable probability to either the
FL$^{2}$S$^{2}$ group or the Magellanic stream group.

The excess clustering at small angular scales seen in the $R_{gc} > 25$
kpc globular cluster$+$DSG sample is due to the crossing points of the
Magellanic DSGs and the following globular
clusters:  Pyxis, Pal~4, NGC~6229, and NGC~7006.  We consider these
four globular clusters to be among the most likely RHB globular
clusters to have been associated with an ancient merger event due to
their tightly clustered orbital pole family crossing points with 
Milky Way DSGs as well as their large Galactocentric radii, which place
them in the outer halo domain of the DSGs.  We further address the possible
association of Pyxis with the Magellanic Clouds in \citet{palma00}.

\subsection{Results of the GCPF Analysis}

The GCPF analysis discussed here is very similar to the ``polar path''
analysis of LB$^{2}$95.  However, we have taken the next logical step
to determine specifically if one subpopulation of the total globular
cluster population is more likely to follow stream orbits than others.
The main conclusion here is that the outer halo, second parameter
globular clusters are much more likely to share orbital poles with the
DSGs of the Milky Way, than the non-second parameter (i.e., Zinn
BHB/MP type) clusters, which are likely to have more randomness in
the distribution of their orbital poles with respect to those of the Milky Way
DSGs.

\begin{figure}
\plotone{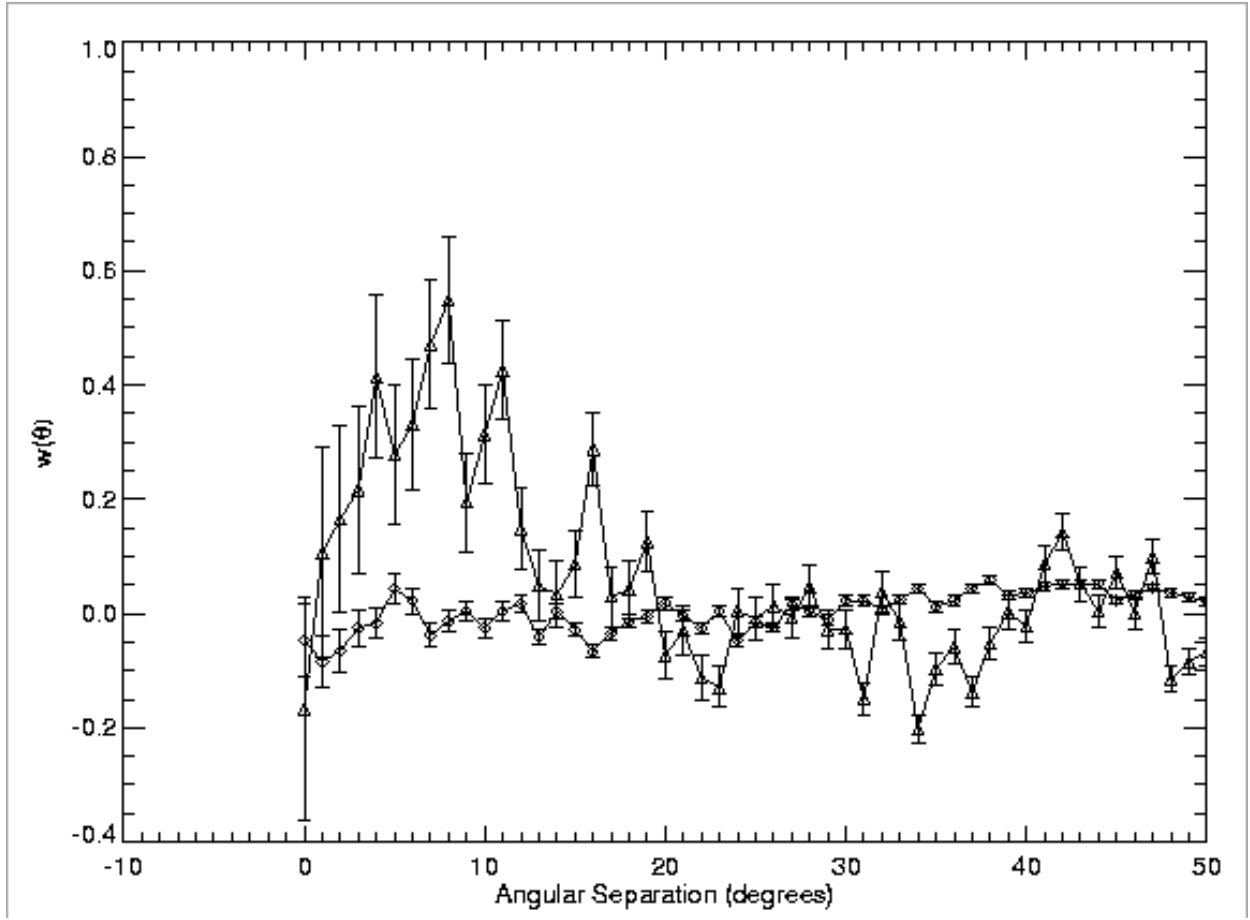}
\caption{Angular two-point correlation function analysis
of the distribution of GCPF crossing points in the $(8 < R_{gc} < 25)$
kpc globular cluster$+$DSG sample (open diamonds) 
versus the $(R_{gc} > 25)$ kpc globular cluster$+$DSG sample (open
triangles).
The clustering amplitude,
$w(\theta)$, is consistent with 0 (or no clustering) for the $8 <
R_{gc} < 25$ kpc$+$DSG sample over the entire range of possible
separations.  In contrast, for $0^{\circ} < \theta \lesssim 15^{\circ}$
the amplitude of the clustering in the $R_{gc} > 25$ kpc$+$DSG sample
is approximately equal to that for the Zinn RHB globular cluster
and DSG sample, with $w(\theta) \sim$0.5.  
This shows that the
clustering seen in the GCPF crossing points at small angular
separations is almost entirely due to the distant outer halo, RHB
globular clusters Pyxis, Pal~4, NGC~6229, and NGC~7006 with the
Magellanic stream and FL$^{2}$S$^{2}$ stream DSGs.
\label{fig:2pt2}}
\end{figure}

It has been previously suggested \citep[e.g.,][]{rp84, lr92, zinn93a}
that the DSGs of the Milky Way may be either the ``fragments'' of
\citet{SZ}, or the remnants of tidally disrupted fragments, and that
perhaps the second parameter globular clusters formed in these
fragments and were later accreted by the Galaxy.  The two-point
correlation function analysis of the GCPFs (Figures \ref{fig:2pt} and
\ref{fig:2pt2}) presented here provides a statistical foundation for
the conclusion that the outer halo second parameter globular clusters
have orbits associated with the Milky Way DSGs.  Neither the Zinn
BHB/MP globular clusters nor those in the metal poor or intermediate
metallicity samples show a positive amplitude in the two-point angular
correlation function analysis of their GCPF crossing points, while the 
Zinn RHB type clusters show a statistically significant 
amplitude. This statistical excess in 
GCPF crossing point clustering for the Zinn RHB globular
clusters$+$DSGs is interpreted here to indicate that there is a
possibility that some of these particular objects are daughter
products of a merger event and 
are currently following similar, stream-like orbits.  However, only
full orbital data (requiring proper motions or perhaps tidal debris
trails that trace the orbits) will bear out this prediction.

Although Zinn RHB globular clusters found at a wide range of $R_{gc}$
contribute to the clump of crossing points seen in Figure
\ref{fig:crosspts}, several $R_{gc} > 25$ kpc RHB globular clusters in
particular contribute to the majority of the clumping seen at small
angular scales typical of tidal debris that we find among the crossing
points.  Of the outer, RHB globular clusters that contribute to the
statistical excess at small angular scales in the clustering of the
GCPF crossing points seen in Figure \ref{fig:2pt2}, several have been
associated with the DSGs in previous studies.  While Pal~4 and NGC~7006
have been proposed members of the Magellanic Plane group for years
(Kunkel \& Demers 1976), and NGC~6229 is included in one of the
LB$^{2}$95 possible streams, the GCPF analysis presented here
associates the recently discovered Pyxis globular cluster with the DSGs
as well.

Currently, several of the DSGs are known to have their own globular
clusters:  Fornax, Sagittarius, and both Magellanic Clouds.
\citet{zinn93b} and, more recently, \citet{smith98} have plotted the
DSG globular clusters in the metallicity vs.\ HB type diagram used by
\citet{zinn93a, zinn96} to separate Galactic globular clusters into the
RHB or BHB/MP types.  There is evidence from this diagram that the
second parameter effect is present in the DSG globular clusters, that
is, they exhibit a spread in HB type at a given metallicity.
\citet{zinn93b} places the LMC/SMC globular clusters in the young halo,
or RHB category.  However, based on a specific age estimate of two
Galactic RHB clusters (NGC 4147 and NGC 4590) that have similar
metallicities and HB types to the Sgr clusters, \citet{smith98} suggest
that the Sgr clusters and the old LMC clusters are more similar to
Galactic ``old halo'' or BHB/MP clusters, even though their location in
the HB type/metallicity diagram is much more like the Galactic
RHB/young halo clusters and at least two of the Sgr clusters are
demonstrably ``young'' from a differential comparison of the morphology
of their stellar sequences to canonical ``old'' clusters
\citep{buon94}.  The \citet{smith98} result may simply reflect the fact
that the second parameter may not be age (a point that prompted Zinn to
switch from the ``young halo'' to ``RHB'' nomenclature).  The Fornax
globular clusters 1, 2, 3, and 5 form a distinct group in the [Fe/H] --
HB diagram:  They have red horizontal branches, however they are more
metal-poor than the Galactic RHB clusters.  Both \citet{zinn93b} and
\citet{smith98} do not consider them RHB (i.e., second parameter)
clusters.  However, recent HST observations of Fornax globular cluster
4 \citep{buon99} have shown it to exhibit a much redder HB than the
other Fornax clusters, even though its metallicity is similar.  Also,
\citet{buon99} find the CMD fiducial lines for Fornax cluster 4 and the
young Galactic RHB cluster Rup 106 are almost identical.  This
observation indicates that there is also a spread in HB type among the
Fornax globular clusters, with at least one Fornax cluster similar to
Galactic RHB clusters.  Since the population of globular clusters found
in the DSGs shows evidence of both the second parameter effect and age
spreads, it is reasonable to posit (as both Zinn 1993b and Majewski
1994 do) that the outer halo, RHB globular clusters, or at
least those found to share orbital poles with DSGs, may
have originated in these galaxies or in other dwarf galaxies that have
already been completely disrupted by the Milky Way.

\subsection{Great Circle Cell Counts}

Figure \ref{fig:crosspts} presents visual evidence that the RHB
globular clusters appear more likely to share similar orbital poles
than do the BHB/MP globular clusters, and Figure \ref{fig:2pt} appears
to confirm this conclusion statistically.  Another way of interpreting
this particular result is to say that more RHB globular clusters are
found distributed along a particular great circle than would be
expected if these objects were randomly distributed on the sky.  Thus,
we might expect the technique of ``Great Circle Cell Counts''
\citep{kvj96} to recover this association and provide
further evidence in support of the non-random alignment on the sky of
the RHB globular clusters and the DSGs.

The technique of Great Circle Cell Counts was designed to search for
stellar debris trails among \textit{large} samples of stars (e.g., from
all-sky surveys) in the halo.  The basis of the technique is to count
sample objects in all possible ``Great Circle Cells'', described by the
two angles that define the direction of the pole of the cell (which are
trivially related to Galactic coordinates), and search for particular
cells that are overdense compared to the background.  Although the
present sample is rather small for this technique, we have nonetheless
calculated Great Circle Cell Counts for the same RHB$+$DSG and
BHB/MP$+$DSG samples presented in \S 5.1.

Using cells of width $2\delta\theta \sim 4.5^{\circ}$, we find that the
RHB$+$DSG sample yields a cell with significance $GC3 = (N_{count} -
\bar{N}) / \sigma_{ran} = 6.2$, where $N_{count}$ is the number of
objects in the cell, $\bar{N}$ is the predicted average number of
objects per cell (which assumes the objects are distributed randomly on
the sky and the number of objects per cell can be described by a
binomial distribution), and $\sigma_{ran}$ is the dispersion around
$\bar{N}$ \citep[see \S 3.2 in][]{kvj96}.  We find that for our various
cluster$+$ DSG samples, most of the cells have $GC3 \leq 3$, so $GC3 >
4$ appears to be the level of marginal significance.  In the
BHB/MP$+$DSG sample, the cell with the highest significance has $GC3 =
4.1$; in order to evaluate the significance of the difference in the
$GC3$ maxima between the BHB/MP$+$DSG and RHB$+$DSG samples, we have
undertaken Monte Carlo simulations of the two samples.

To construct samples for the Monte Carlo test, we used the same
randomization algorithm applied in the two point correlation function
analysis (presented in the Appendix) to create 1000 random datasets
from the RHB$+$DSG sample and from the BHB/MP$+$DSG sample.  We
analyzed these randomized datasets with the GC3 technique to estimate
the statistical significance of the clump found in the RHB$+$DSG
sample.  After applying cylindrical randomization (see Appendix), 1.3\%
of the 1000 randomized RHB$+$DSG datasets had a cell with $GC3 \geq
6.2$.  However, among the 1000 randomized BHB/MP$+$DSG datasets, 73.4\%
had at least one cell with $GC3 \geq 4.1$.  After applying spherical
randomization, the percentages in both cases go down a bit (0.3\%
RHB$+$DSG datasets have $GC3 \geq 6.2$ and 69.5\% BHB/MP$+$DSG datasets
have $GC3 \geq 4.1$), however this may reflect a selection bias:  Due
to the Zone of Avoidance, great circle cells with poles near the
Galactic pole are not as likely to be found to have large values of
$GC3$.  While the cylindrical randomization preserves the $Z$
distribution of the satellites (and the inherent likelihood function
for significance as a function of Galactic latitude), the spherical
randomization algorithm dilutes the likelihood for significance among
great circle cells inclined to the Zone of Avoidance by making
\textit{all} inclinations equally likely to be found significant.
Thus, due to its preservation of the bias resulting from the Zone of
Avoidance, the cylindrically randomized test data offer a more fair
comparison to the real data than do the spherically randomized
datasets.

In addition to offering a means to test the \textit{statistical
significance} of the previously identified RHB$+$DSG clump, the Great
Circle Cell Count technique also allows an independent check of the
\textit{specific} clump by returning the pole of the cell with the
maximum number of counts.  As expected based on the GCPF crossing point
analysis, the pole of the cell in the RHB$+$DSG sample with the highest
statistical significance has $(l,b) = (175,-22)^{\circ}$; this cell
corresponds to the clump of crossing points identified previously,
which we estimated to be centered near $(165,-25)^{\circ}$.

To summarize, our analysis of the globular cluster and dwarf satellite
galaxy populations indicates that the RHB globular clusters exhibit a
non-random spatial distribution that may be attributed to their
following similar orbits, while the BHB/MP globular clusters appear to
be randomly distributed around the galaxy.  We have demonstrated this
point in two ways: First, the two point angular correlation function
test showed the clumping among the GCPF crossing points of the RHB$+$DSG
sample to have a much higher statistical significance than for the
BHB/MP$+$DSG sample.  Second, by counting objects in all possible great
circle cells, we have identified a great circle cell in the
RHB$+$DSG sample that contains more objects than expected for a randomly
distributed sample, yet we find no similar excess in any great
circle cell in the
BHB/MP$+$DSG sample.

\section{Arc Segment Pole Family Analysis}

Although the GCPF technique provides a statistical means for
identifying samples of globular clusters that have a significant
probability of following orbits similar to the Milky Way 
DSGs, the ASPF technique may allow us to identify individual captured
clusters directly.  In this section, we describe the calculation of the
ASPFs for the sample of halo objects with measured proper motions and
present one method of codifying the significance of the 
clustering of the ASPFs for those objects with similar ASPFs.

\subsection{Pole Families From Independent Proper Motion Measurements}

We calculate ASPFs for each object and each independent proper motion
measurement (Table \ref{pmtab}).  In cases where there are multiple
proper motion measurements, we also calculated an ASPF from the
unweighted average of these measurements.  Often the discrepancies from
measurement to measurement for the same object were large, and it was
not clear that taking an average of two or more widely different
measurements gave a more accurate result.  We therefore selected what
we considered to be the most precise proper motions from among the
various independent measurements, using somewhat subjective criteria.
In most cases, this amounted to adopting the measurement with the
smallest quoted error.  The largest discrepancies between independent
measurements seem to exist when comparing proper motions measured from
Schmidt plates and those measured from finer scale plate material and
this leads us to suspect problems with the former.  Therefore, in some
cases when a proper motion derived from Schmidt plates had the smallest
quoted error (M3, M5, M15, Pal~5), we
did not choose to use it if the value was highly discrepant from other
measurements.  Moreover, due to the often large discrepancies between
the {\it Hipparcos}-calibrated proper motions of Odenkirchen et
al.\ (1997) and multiple previous determinations (e.g., in the cases
of 47 Tuc, M3, M5,
M92), we only chose to use {\it Hipparcos}-calibrated
measurements if no other proper motion
was available.  In most cases our choice of proper motion had little
impact on our ASPFs or conclusions; however, in four cases where the
differences were significant, (Pal~5, M3,
M5, and M15) we did calculate and analyze ``alternate''
ASPFs;
these cases are considered below.

We have followed the recommendation of Dinescu et al.\ (1999b) on the
proper motion of NGC~362 and adopt the Tucholke (1992b) proper motion
of this object, which is calculated relative to SMC stars.  However,
Dinescu et al.\ (1999b) derives an absolute proper motion for NGC~362 by
correcting for the SMC's proper motion using the Kroupa \& Bastian
(1997) measurement.  The corrected ASPF has a center more than
50$^{\circ}$ from the center of the uncorrected proper motion's pole
family.

\subsection{The Distribution of ASPFs}

A simple hierarchical clustering algorithm \citep{murtheck} was used to
search for statistically significant groups in the distribution of
ASPFs.  Since the arc segments are not points, the ``distance'' between
two arcs is not well-defined.  Therefore, the angular separation along
a great circle between each point along the one arc and each point on
the other arc was measured.  The pair of points that gave the {\em
minimum} angular separation between the arcs was selected, and we
defined this minimum separation as the distance between the two arcs.
Using the angular separation between arc {\em centers} as the distance
measure was also investigated, and no significant differences
in the results were obtained.

The steps in the cluster analysis algorithm are as follows:

\begin{enumerate}

\item Construct a matrix containing the minimum angular separation between all ASPFs
in the sample.

\item Identify the pair of ASPFs $A$ and $B$ separated by the smallest 
distance in the matrix.

\item Replace ASPFs $A$ and $B$ in the matrix with $A\cup B$.

\item Update the matrix by deleting ASPF $B$, and replacing the
distances to all other matrix ASPFs from ASPF $A$ with those of $A\cup
B$ where: \\
 $d_{A\cup B,C} = \alpha_{A}d_{A,C}+\alpha_{B}d_{B,C}+\beta
d_{A,B}+\gamma |d_{A,C}-d_{B,C}|$.

\item Repeat from step 1 until all ASPFs in the sample have been agglomerated.

\end{enumerate}

The group selection is determined by the choice of the coefficients in
the updating formula (step three).  If one sets $\alpha_{A}=0.5$,
$\alpha_{B}=0.5$, $\beta=0$, and $\gamma=-0.5$, then the algorithm is a
single linkage, or nearest neighbor algorithm.  With these
coefficients, $d_{A\cup B,C} = \min(d_{AC},d_{BC})$, and points get
pulled into a group if they lie near any of the objects which make up
the group.  However, this method has the disadvantage of tending to
identify long, stringy clusters.  Since we are trying to identify
clusters among arc segments rather than discrete points, 
this ``chaining'' problem is exacerbated
if we apply the single linkage algorithm to the ASPFs.  If one has two
arc segments nearly end to end, the algorithm finds that they have a
small separation and will link the two of them with a very small
distance even if the centers of the arcs are widely separated.  If
this pair intersects another such pair, you have a long, narrow
``group'' of pole families that may lie on an arc nearly 180 degrees in
length.  Since the true orbital pole of an object lies only at one
point along an arc, clearly, a set of four arcs end to end would not
constitute what one would call a true group of orbital poles.

We therefore rely on a centroid or average linkage technique to avoid
identifying spurious ``chain''-type groups.  If one sets
$\alpha_{A}=|A|/(|A|+|B|)$ (here $|A|$ denotes the number of ASPFs in
cluster $A$) and $\alpha_{B}=|B|/(|A|+|B|)$,
$\beta=-(|A||B|)/(|A|+|B|)^{2}$, and $\gamma=0$, then the updated
distance $d_{A\cup B,C} = $ the distance from the centroid of the group
containing $A$ and $B$.  With this method we are biased towards finding
small, centrally concentrated groups rather than chains.

\begin{figure}
\epsscale{0.6}
\plotone{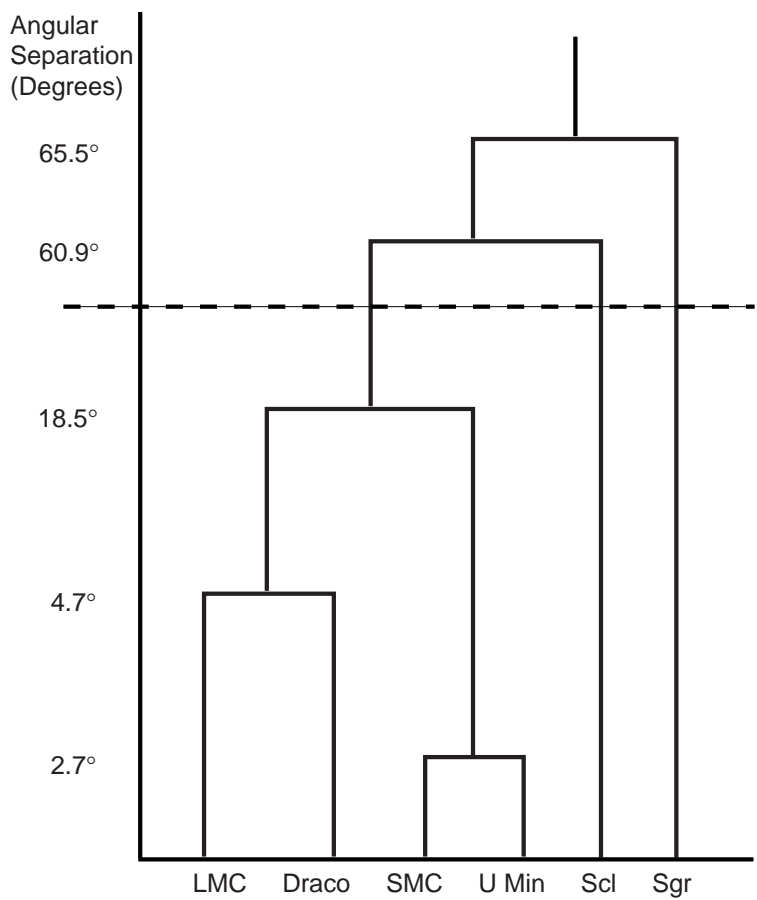}
\caption{Dendrogram representation of the output from the centroid
cluster analysis algorithm. This dendrogram has been constructed using
the output for the 6 Milky Way satellite galaxies with known space
motions, as listed in Table \ref{dendtab}.  Objects (or groups of objects)
connected by a horizontal line have been agglomerated.  The lowest
horizontal line indicates the first pair of objects to be agglomerated,
and each subsequent agglomeration is indicated by a successively higher
horizontal line.  Listed along the ordinate are the angular distances
separating each pair.  Note the large (nearly a factor of four) 
jump in the angular
separation between Sculptor and the Magellanic stream satellite
group.  The dotted line indicates the partition constructed by
identifying the first large jump in distance from rank to rank, groups
below this line may be considered real, while those above the line are
probably spurious.
\label{fig:dend}}
\end{figure}

The output of the algorithm is a series of agglomerations and the value
of the distance at which the pair was agglomerated.  For example, Table
\ref{dendtab} shows the output for the centroid algorithm for the six
DSGs with known space motions.  One can represent
the output of this algorithm graphically in a dendrogram; Figure
\ref{fig:dend} is an example of a dendrogram drawn using the centroid
linkage data from Table \ref{dendtab}. The simplest interpretation of
the algorithm's output is obtained by partitioning the output between
two ranks of the hierarchy.  All groups found below the partition can
be considered ``real'' and those of higher rank not.  This is
represented in the dendrogram by a horizontal line between the last
``real'' rank and those with larger dissimilarities.  

There are several
methods for selecting the partition.  If one has no physical intuition
for the size scale that separates real groups from spurious
associations, the partition can be drawn between the two ranks that
exhibit the first large jump in the dissimilarity measure from the one
rank to the next.  The alternative is to set a predetermined limit
defined using {\it a priori} information about the sample.

As discussed in \S3.2, we expect daughter products of a common merger
event to have had their orbital poles spread from their initial
alignment.  If we can estimate this spread, we can use it as a
constraint on the partition that separates real groups from spurious
associations.  Several scales can be taken from previous work on
streams in the halo.  LB$^{2}$95 suggest that one could, for example,
use the angle that the tidal radius of Fornax (at the time, the only 
known DSG with its own population of globular clusters) subtends.
LB$^{2}$95 quote a current
value is 1$^{\circ}$ for Fornax, but they suggest that the proper scale may
be up to 4$^{\circ}$ depending on how close Fornax is to its
perigalacticon.  The alternative scale suggested in LB$^{2}$95 is that
of the spread in angular size of the gas contained in the Magellanic
Stream; they suggest adopting either the 5$^{\circ}$ width of the
Stream or the separation of the LMC and SMC as projected along the
Stream, or about 15$^{\circ}$.  The tidal scale, $f$, (see eq.
[\ref{de}]) for the Milky Way DSGs ranges from
$\sim$1$^{\circ}$ to $\sim$16$^{\circ}$.  We also consider the scale
adopted by \citet{srm94} who calculated the probability that in a
random distribution a globular cluster would lie closer to the
FL$^{2}$S$^{2}$ Plane than it does at the current epoch.  The angle
used to separate globular clusters into groups anti-correlated
and correlated with the FL$^{2}$S$^{2}$ Plane varies between
20$^{\circ}$ and 30$^{\circ}$, depending on the globular cluster's
$R_{gc}$.

As there is no consensus on the exact angular scale that divides
correlated orbital planes from uncorrelated ones, and since the angular
separation between arc segments can be measured in more than one way,
we do not rely solely on a predefined angular size as our partition.
Instead, we adopt 20$^{\circ}$ as an absolute upper limit on the
separation between poles in a group, but typically use a large jump in
the dissimilarity between ranks below this upper limit as a more
conservative partition.

As an example, using the guidelines described in the preceding
paragraph, we could partition the data in Table \ref{dendtab} between
ranks 3 and 4, where the angular separation jumps from 18.5$^{\circ}$
to 60.9$^{\circ}$.  Any agglomerations found in rank 3 and below have
the most closely aligned angular momentum vectors and define the groups
we consider to have the highest probability of being dynamically
related.  Based on the groups below the partition in Figure
\ref{fig:dend}, one concludes that Sculptor and Sagittarius are not
likely associated with the Magellanic stream group of LMC, SMC, Draco,
and Ursa Minor.

\begin{figure}
\epsscale{1.0}
\plotone{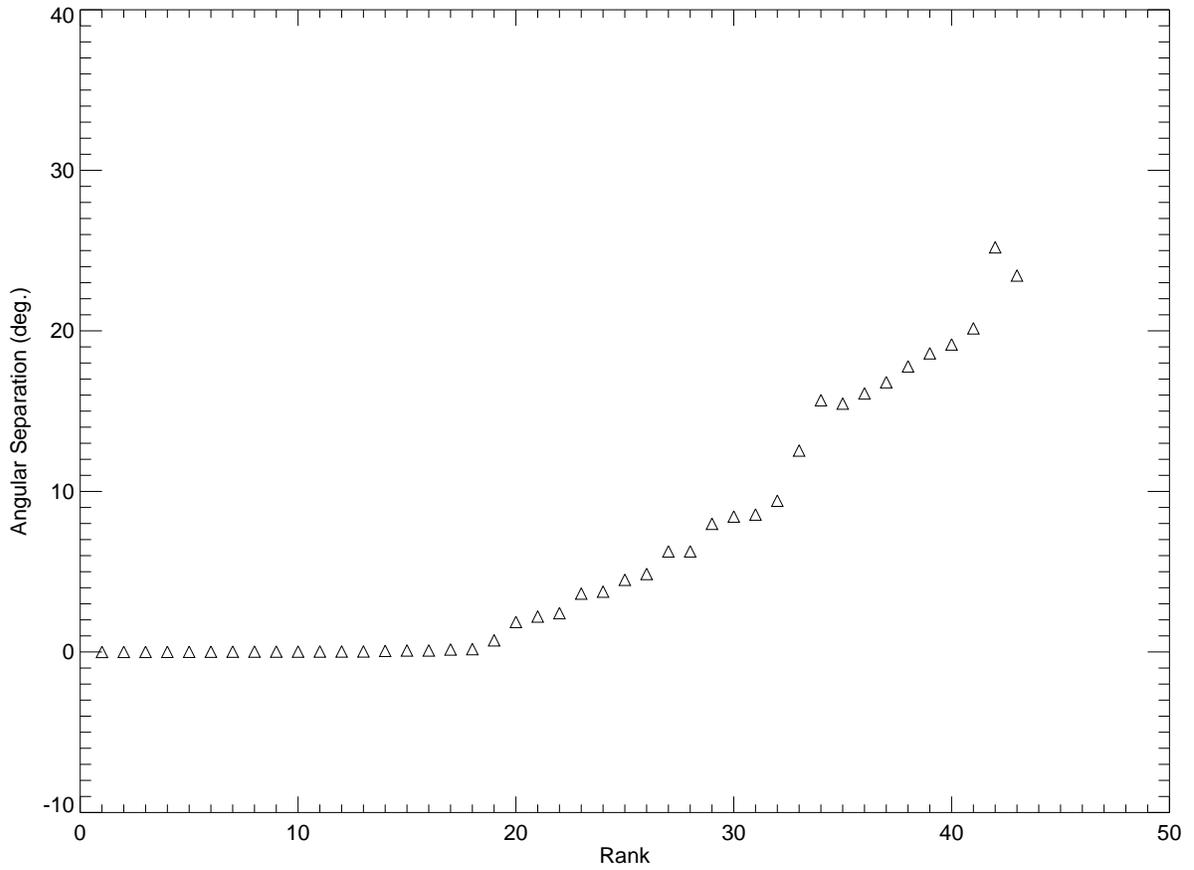}
\caption{The angular separation between ASPFs at each rank in the agglomeration
algorithm.  Since the expected angular separation between ASPFs is not
well-constrained, the first large jump in angular separation between
successive ranks was adopted as the partition between potentially real
groups and those having lower probabilities of being true dynamical
groups.
\label{fig:agglom}}
\end{figure}

\begin{figure}
\plotone{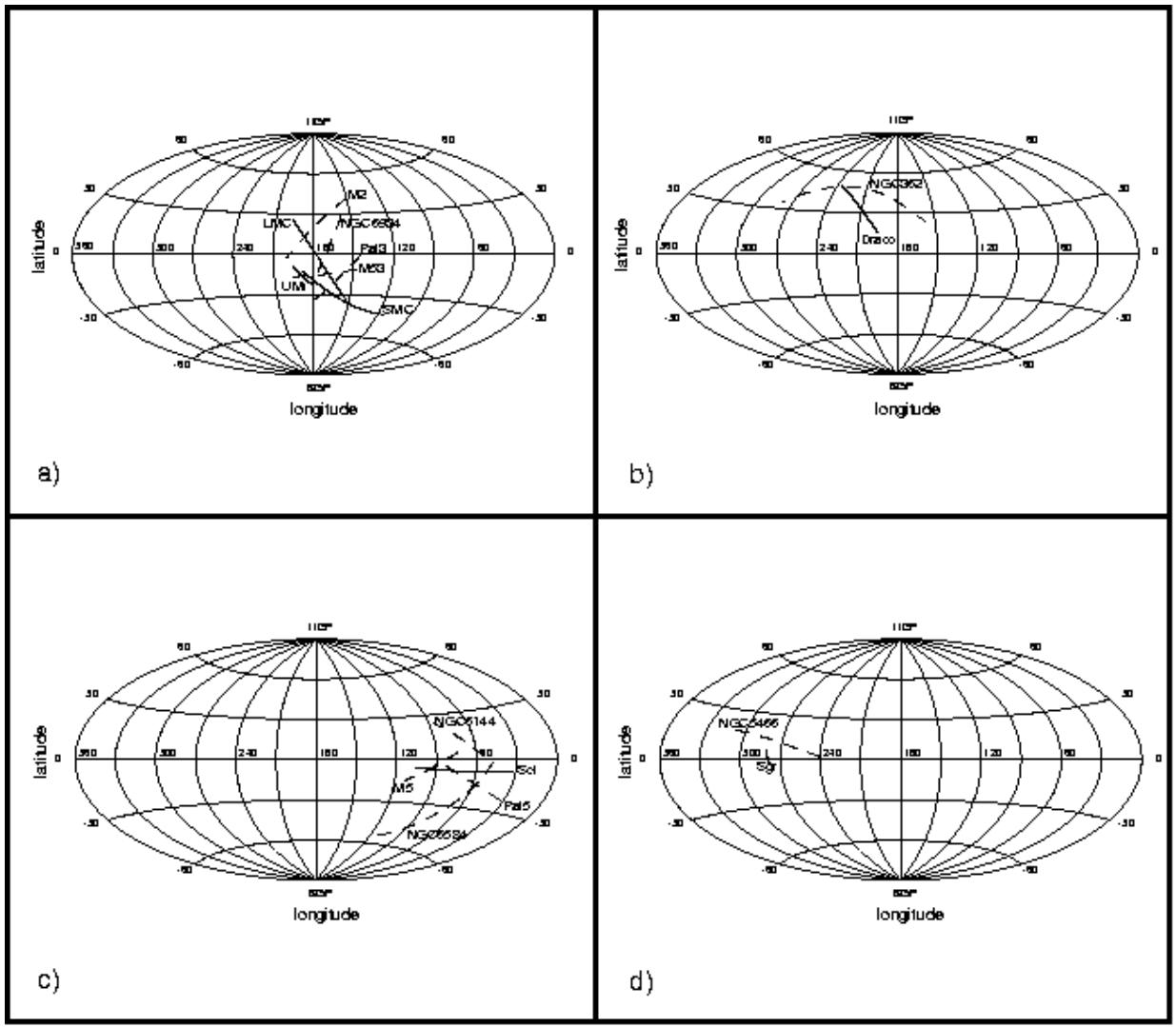}
\caption{Each panel shows the ASPFs of objects we find to be grouped,
implying that they have nearly coplanar orbits.  Dashed lines represent
the possible orbital poles of globular clusters and solid lines
represent those of DSGs.  a) The Magellanic stream Group, which
includes the LMC, SMC, and Ursa Minor dwarf galaxies, is found in our
analysis to also include the pole families of the globular clusters M2,
M53, NGC~6934 and Pal~3.  b)  Draco is usually included in the
Magellanic stream Group \citep[e.g.,][]{lb82}, but its ASPF is fairly
distant from the nexus of the ASPFs of the LMC, SMC, and Ursa Minor.
We group NGC~362 with Draco based on the proximity of their ASPFs.  c)
The ASPF for the Sculptor dwarf is found to cluster with the pole
family of the globular cluster M5, NGC~6144, NGC~6584 and Pal~5.
However, an alternative proper motion for Pal~5 places its ASPF far
from that of Sculptor.  d) The Sagittarius dwarf ASPF is relatively
isolated, however, the ASPF of NGC~5466 is grouped with Sagittarius by
our cluster analysis algorithm.
\label{fig:4pan}}
\end{figure}

We performed a statistical cluster analysis on the entire sample of 41
globular clusters and six DSGs with published space
motions (and therefore ASPFs).  A partition between ranks 32 and 33,
where the angular separation between arcs jumped by 33\%, was adopted.
Figure \ref{fig:agglom} presents a plot of the angular separation
measured at each rank, and illustrates the ``jump'' that was selected
as the partition.  Below our adopted partitions we find the following
objects to have grouped ASPFs (see Figure \ref{fig:4pan}):

\begin{description}

\item[Magellanic stream Group]  We find five globular clusters that
have pole families aligned with the ASPFs of the Magellanic
stream group of galaxies (LMC, SMC, Draco, and Ursa Minor).  Below our
partition, these galaxies are isolated into three separate subgroups that
also contain various globular clusters.  The first contains the pole
families of the LMC, M2, and NGC~6934.  The second contains
Draco and NGC~362.   The final subgroup contains SMC, Ursa Minor, Pal~3,
and M53.  Just prior to the partition, the LMC and SMC subgroups
merged.  

\item[Sagittarius Group]  The ASPF of Sagittarius is found to 
be aligned with the ASPF of only one globular cluster with a 
presently known orbit, NGC~5466.

\item[Sculptor Group]  The Sculptor ASPF is intersected by the pole
families of NGC~6584, Pal~5, M5, and NGC~6144.  Only the ASPF
generated for Pal~5 using the Scholz et al.\ (1998) proper motion 
intersects that of Sculptor.  As noted in \S6.1, the Scholz et
al.\ (1998) Pal~5 proper motion was measured from Schmidt plates, and
is widely discrepant with the Cudworth et al.\ (2000) measure.  The
ASPF for Pal~5 generated from the Cudworth et al.\ (2000) proper motion
lies in a different part of the sky, and is unassociated with
Sculptor.

\end{description}

\subsection{Dynamical Groups}

The groups described in the previous section are selected purely 
on the basis of current estimates of the locations of their orbital poles
(proximity of ASPFs).  However, before one can claim that the above
groups are indeed dynamical associations leftover from a past merger
event, one must also compare the size and shape (equivalently, energy
and angular momentum) of the member orbits.  We now evaluate the
likelihood that the above groups are true dynamical groups from this
standpoint.

If one assumes that the orbits of the daughter objects of a tidal
disruption of a parent object retain approximately the same specific
angular momentum and specific energy as their parent, then we can sort
true dynamical groups from chance alignments of ASPFs using the
$z$-component of the angular momentum and the orbital energy, which are
conserved exactly in static, oblate potentials which is likely
applicable to the Milky Way \citep[cf.][regarding the shape of the halo]
{lh94}.  Since the calculation
of $E$ requires assumptions about the shape of the potential, $L_{z}$
is likely to be a better discriminant between dynamical groups and
spurious associations than is energy.  However, the errors in $L_{z}$ are
in many cases large due to the imprecision of the proper motions.
Mindful of these shortcomings, we can nevertheless use $L_{z}$ and $E$
to rule out spurious associations in cases where the values of these
two quantities differ grossly for objects with aligned ASPFs.  We also
consider the total angular momentum, $\vec{L}$, which is conserved in
spherical potentials and will be nearly conserved in potentials that
are almost spherical.  Therefore, even if the potential is 
oblate, associated objects at similar orbital phase should have similar
$|L|$, and will be identifiable by having similar
$\vec{L}$ {\em directions} (i.e., their ASPFs are nearly aligned).
In Table \ref{orbtab} we list $|L|$,
$L_{z}$, and $E$ \citep[calculated using the potential in][]{kvj95} for
the objects we list above as having aligned ASPFs.

In simulations of the disruption of satellites, Johnston (1998) found
that the majority of the debris maintains orbits with energy within
$\pm 3 dE$ (defined in eq. [\ref{de}]).  To aid with comparison to this
tidal scale, Table \ref{orbtab} lists values of $dE$, $dL_{\rm tot} =
f|L|$, and $dL_{\rm z} = fL_{\rm z}$ calculated from the mass of the
largest object in the group.  This tidal scale estimate may be a
conservative limit under the assumption that the most massive group
member has shed mass to create the other group members.  Note that
these scales are often smaller than the observational errors associated
with the derived quantities and hence it is difficult at this point to
make a meaningful comparison of the dispersion among energy and angular
momentum within each group to the tidal scale.

Despite large error bars, the results listed in Table \ref{orbtab} do
allow us to rule out several ASPF associations.  The Sculptor Group
globular clusters show little overlap among their $L_{z}$ or $E$ values
with Sculptor, indicating a low probability that this is a valid
dynamical association.  Magellanic Group 2, which consists of Draco and
NGC~362, have very different orbits and are clearly unassociated.  The
space velocity for Draco gives it an $E$ and an $L_{z}$ much larger
than any of the other Milky Way satellite galaxies, and in the Johnston
et al.\ (1995) potential its orbit does not return it to the Milky Way
within a Hubble time.  The Scholz \& Irwin (1994) proper motion for
Draco is very similar to their proper motion for Ursa Minor (see Table
\ref{pmtab}), and in both cases is fairly large.  On the other hand,
the Schweitzer et al.\ (1997) Ursa Minor proper motion, which was
measured from plate material with a finer plate scale and fewer
distortions, is about an order of magnitude smaller in each component
than that of Scholz \& Irwin (1994).  This discrepancy for Ursa Minor
suggests, by analogy, that the true proper motion for Draco may be
smaller than the one used here.  The exclusion of Draco from the
Magellanic stream group of objects based on its current proper motion
may therefore be a premature conclusion.

\begin{figure}
\plotone{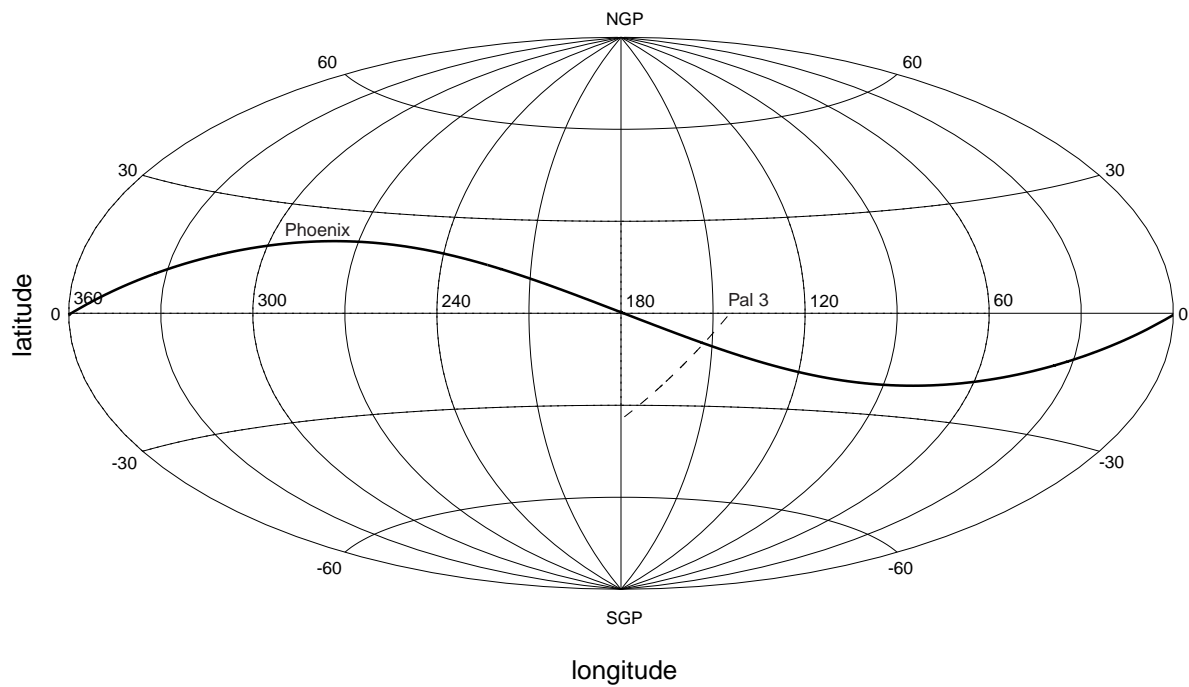}
\caption{A potential association between the globular cluster Pal~3 and
the Phoenix dwarf.  Dinescu et al.\ (1999b) finds that integrations of
Pal~3's orbit show that it is near perigalacticon currently, and that
it will reach more than 400 kpc from the Milky Way's center when it
reaches apogalacticon.  The Phoenix dwarf has an $R_{gc}$ of 445$\pm30$
kpc, and may be on a nearly coplanar orbit to that of Pal~3. 
\label{fig:phepal}}
\end{figure}

Although we find the ASPF of the globular cluster Pal~3 to group with
those of SMC and Ursa Minor, a more likely association may be Pal~3 and
the Phoenix dwarf.  Clearly, the current values of the orbital parameters
for Pal~3 are not very similar to the Magellanic Group 3 DSGs.
As Figure \ref{fig:phepal} illustrates, the ASPF of Pal~3 also
crosses the GCPF of the Phoenix dwarf.  Moreover, the orbital integrations
of Dinescu et al.\ (1999b) give an apogalacticon for Pal~3 of $>410$
kpc, a value that is close to the current distance of the Phoenix dwarf
of $\sim$445 kpc.

The only DSGs that appear to share similar orbital parameters
are Ursa Minor and the Small Magellanic Cloud.  Based on the current
best estimates of their space velocities, the LMC may be associated
with Ursa Minor and the SMC.  We calculated the ASPF for the LMC using
the Jones et al.\ (1994) proper motion, however their reduction of this
proper motion to an absolute reference frame was complicated by the
unknown amount of rotation of the stars around the center of the LMC.
If we instead calculate the ASPF using the Jones et al.\ (1994) LMC
proper motion corrected for the estimated amount of rotation in the
field or with the Kroupa \& Bastian (1997) LMC proper motion, we find
that the ASPF in either of these cases lies almost entirely in the
northern Galactic hemisphere, indicating that for the LMC, $L_{z}$ is
positive, and opposite that of the SMC and Ursa Minor.  However, since
these objects are all on nearly polar orbits and are in the outer halo
where the potential is more nearly spherical, the difference in the
sign of $L_{z}$ may not be meaningful.

Several of the globular clusters listed in Table \ref{orbtab} {\em may}
be on orbits similar to the DSGs listed in the same group.  The error
bars on $L_{\rm z}$ for M2, NGC~6934 and the LMC overlap
each other within $1\sigma$, as do those of M53 and the SMC,
and Sagittarius and NGC~5466.  A similar overlap is seen in the values
of $E$ for these pairs of DSGs and globular clusters.  These results
await improved proper motion determinations for verification.  The
proposed measurement accuracy for proper motions measured by the Space
Interferometry Mission will be in excess of what is needed to verify
these potential associations.

In \S5.1, eleven globular clusters were selected that may belong to the
Magellanic stream or Fornax--Leo--Sculptor stream because their GCPFs
intersect near the nexus of GCPF intersections of the DSGs.  Of this
group, there are proper motions for four of them, Pal~3, M3, M68, and
NGC~4147.  Pal~3 is the only one of the four that has an ASPF that
places its orbital pole directly within this cluster of crossing
points.  Below the partition of $9.5^{\circ}$ that was adopted during
the cluster analysis of the ASPFs, the other three (M3, M68, and NGC~4147)
were not agglomerated into the Magellanic stream Group.  However,
the ASPFs of these objects appear by eye (Figure \ref{fig:pyxisetc}) to
lie remarkably close to the center of the Magellanic stream Group, and
it is possible that more precise proper motions (e.g., from SIM) for
M3, which was measured from Schmidt plates, and NGC~4147, which has
large measurement errors, may force reconsideration of the inclusion
of these clusters with the Magellanic stream Group.  The proper motion of Pal~3 has a
large uncertainty as well \citep[$0.33\pm0.23$,$0.30\pm0.31$
mas/yr;][]{srmkmc93}, and was ruled out as an associate of the
Magellanic stream Group based on its large angular momentum and
energy.   Should a revised proper motion show that the true velocity of
Pal~3 lies at the lower end of the range suggested by the large error
bar, Pal~3 may also have to be reconsidered as a member of the
Magellanic stream Group.  The ASPFs of the three globular clusters, M3,
M68, and NGC~4147 also appear unassociated with that of Sculptor, but we
can not rule out an association with other FL$^{2}$S$^{2}$ Stream
galaxies since Fornax, Leo I, Leo II, and Sextans do not yet have
measured proper motions.

\subsection{ASPF Analysis and Zinn RHB Globular Clusters}

Of the 41 globular clusters with known space motions and tested for
common orbits with the DSGs, only four clusters are found to show
potential dynamical associations.  This is not very surprising,
however, because of these 41 globular clusters, 21 have $R_{gc} < 8$
kpc, and all except Pal~3 and Pal~13 have $R_{gc} < 25$ kpc.  The inner
halo is predominantly populated by the Zinn BHB/MP globular clusters
that are likely to have originated during the ELS collapse phase of the
inner Milky Way's formation \citep{zinn93a}.  Unfortunately, few outer
halo RHB clusters have known orbits, and yet it is these that are more
likely to have been accreted (see discussion in \S1).

Three globular clusters that may be dynamically associated with the
Magellanic stream or Sagittarius DSGs based on our ASPF analysis are
M2, M53, and NGC~5466.  These three are all Zinn BHB/MP type, contrary
to expectations.  Since the coincidences among the dynamical quantities
are not perfect, one could argue that these three are simply random
alignments that would arise in any sample of 41 globular clusters
analyzed in the same way.  However, there are 11 Zinn RHB type globular
clusters among the 41 globular clusters analyzed with the ASPF
technique, and it is curious that only one of these eleven (NGC~6934 is
in Magellanic Group 1) is found to be dynamically associated with the
DSGs using the ASPF technique.  It is possible, in the end, that the
present sample of objects with which the ASPF technique may be applied
is still too confined to the inner Galaxy, where signs of dynamical
association are most likely to be erased.

The following is a summary of our knowledge of the orbits of the RHB globular 
clusters:

\begin{enumerate}

\item Pal~3, Pal~5 \citep[using the proper motion of][]{scholz98}, 
and NGC~362 were grouped with the SMC, Sculptor, and Draco
respectively, but the orbital parameters do not support these
as dynamical associations.  However, Pal~3 may be associated with
Phoenix, and the \citet{kmc01p5} proper motion associates Pal~5
instead with Sgr.

\item \citet{dd00} associate Pal~12 with Sgr, but we do not find
the poles of these two objects to be presently well-aligned.  
However, this may not be surprising given that the postulated disruption
occured $\sim$1.7 Gyr ago and the pole of Pal~12 may have precessed
or otherwise drifted away from the pole of Sgr. 

\item \citep{pal13} point out that Pal~13 does not have an orbit
similar to Sgr, nor does it appear to be associated with any of the
other DSGs.  Pal~13 was not agglomerated into any of our ASPF groups;
however, the Pal~13 ASPF does intersect the GCPFs of Leo I and
Sextans.  The apogalacticon of Pal~13 is 81 kpc \citep{pal13},
suggesting a more likely association with Sextans ($R_{gc} \sim$85
kpc), than Leo I ($R_{gc} \sim$250 kpc).

\item NGC~4147, M68, and M3 have ASPFs that lie within
$\sim$26$^{\circ}$ of the center of Magellanic Group 3 (the SMC, Ursa
Minor, Pal~3, and NGC~5024).   As proposed in the case of Pal~12 above,
differential precession may have caused this spread among the poles of
these clusters. But the orbital parameters of these objects are not
widely discrepant from those of the Magellanic group:  NGC~4147 has
orbital parameters similar to those of the SMC and Ursa Minor, and
NGC~4590 has $L_{z}$ almost identical to NGC~4147 (-25$\times 10^{2}$
and -24 $\times 10^{2}$ kpc km/sec respectively), although the
magnitude of the orbital energy of NGC~4590 is unlike that of NGC~4147,
the SMC, or Ursa Minor.

\end{enumerate}

\section{Summary and Conclusions}

We have undertaken several analyses to ascertain whether dynamical
families spawned from the break up of parent objects are identifiable
today under specific assumptions, i.e., that the objects retain similar
orbital poles over long timescales.  Our case study of the Sgr system
reveals that several globular clusters with properties similar to those
previously identified Sgr globular clusters may share the orbital pole
of Sgr.  Although none of the globular clusters are definitely
classified by us as Sgr clusters, Pal~5, M53, M5, and NGC~6356 may be
considered candidate Sgr clusters worthy of further investigation.

We have constructed Great Circle Pole Families (GCPFs) for all Milky
Way globular clusters and DSGs by finding the plane that contains all
possible normals to the radius vector of each object.  To identify
possible dynamical associations, we select clumps in the distribution
of GCPF crossing points.  We find a quite large amplitude of clustering
in the crossing point distribution among $R_{gc} > 8$ kpc second
parameter (RHB, or ``Young'' Halo) globular clusters.  In fact, most of
the clustering among the globular cluster GCPF crossing points at
angular scales $<15^{\circ}$ comes from second parameter globular
clusters with $R_{gc} > 25$ kpc.  We conclude that those distant RHB
globular clusters whose GCPFs create the excess of GCPF crossing points
are those most likely associated with the Magellanic stream or
FL$^{2}$S$^{2}$ \citep{lb82} DSGs.  The possible member clusters are
Pyxis, Pal~4, NGC~6229 and NGC~7006.  Since the crossing points of
these four globular clusters and the DSGs lie primarily near the nexus
of GCPF crossing points for the Magellanic stream group, it is more
likely that these are Magellanic stream members than FL$^{2}$S$^{2}$
stream members.  The GCPF analysis gives statistical support to the
association of outer halo second parameter globular clusters and the
DSGs.  This strengthening of previous suggestions
\citep[e.g.,][]{srm94} that second parameter globular clusters share
nearly coplanar orbits with DSGs promotes the case for their mutual
origin in past accretion events in the Milky Way's outer halo.  The
GCPF results suggest to us that whether or not the physical cause of
the second parameter is age, a common origin of second parameter
globular clusters with the DSGs may explain the source of the second
parameter effect.  

It has been shown that the few DSGs that have their own populations of
globular clusters (Sgr, Fornax, and both Magellanic Clouds) contain RHB
globular clusters, BHB/MP globular clusters, and a few that seem to fit
neither category.  In the metallicity vs.\ HB type diagram, the old
clusters in the Magellanic Clouds and most of the Sgr clusters are
similar to Galactic RHB clusters, while most of the Fornax clusters are
unlike any Galactic clusters, though they do have second parameter-like
behavior (they have red HBs, but are more metal-poor than any of the
Galactic RHB clusters).  The globular cluster system in Fornax seems to
exhibit a second parameter dichotomy. It is worth noting that such
second parameter dichotomies are also found among the
\textit{non-cluster} stars in some DSGs, for example in the Sculptor
\citep{srm99b, hk99} and And I dwarf spheroidals \citep{DaC96}.  On the
whole, Galactic DSGs contain a larger fraction of second parameter-like
globular clusters than does the Milky Way; therefore it is not
unreasonable to suggest that RHB clusters are more predominantly
associated with DSGs and that perhaps \textit{all} RHB clusters
ultimately may have derived from a DSG.  Given the association of RHB
clusters and RHB stellar populations with DSGs, it would appear that
something particular to DSG environments promotes second parameter
expression.

Using velocity data culled from various sources in the literature, we
calculated the range of possible directions for the vectors defining
the orbital poles of Milky Way satellites: 41 Galactic globular
clusters and 6 DSGs.  Our application of cluster analysis algorithms
allowed us to identify which of the globular clusters were most likely
to share a common orbital plane with the various DSGs. Unfortunately,
the small number of globular clusters with proper motions is dominated
by non-RHB clusters, and so is short on examples of the type of
globular cluster we expect to be dynamically associated with the DSGs
based on the results of our GCPF analysis.  However, we do
find a few potential dynamical families.  The orbital parameters
derived from the space motions of the SMC and Ursa Minor are very
similar, and the LMC may also be on an orbit rather closely matching
those of these two galaxies.  Among our other tentative globular
cluster/DSG associations, there is some evidence that the dynamics of
the LMC, NGC~6934, and M2 are similar, and we find that M53 is on an
orbit like that of the SMC and Ursa Minor.  Finally, the angular
momentum and energy of NGC~5466 are within $1\sigma$ of those for the
Sagittarius dwarf.  Three of the just-named globular clusters are Zinn BHB/MP
type (all except NGC~6934), and these associations may therefore be
random alignments since previous studies and our GCPF analysis suggest
that the Zinn BHB/MP globular clusters are likely to have originated in the
collapse phase of the Galaxy's formation and not by accretion from a DSG.

The GCPF analysis suggests that the globular clusters Pal~3, 
M3, M68, and NGC~4147 are potentially associated with the
Magellanic stream Group.  Although the ASPFs of these objects were not
agglomerated into the Magellanic stream Group in our statistical
analysis, they all lie within $\sim$10-30$^{\circ}$ of the orbital
poles of Ursa Minor and the SMC, and uncertainties in the proper
motions of Pal~3, M3, and NGC~4147 are large enough that we
can not rule out their inclusion with this group of DSGs.  Also, these
objects are at relatively small $R_{gc}$, where precession will cause
orbital poles to spread more rapidly than they would at larger
$R_{gc}$.  Therefore, full orbital integrations calculated over a
Hubble time may reveal that at earlier times the orbital poles of these
globular clusters may have been closer to the poles of the Milky Way
DSGs than they are now.  Again, more precise proper motions are
necessary to accurately reconstruct the orbital paths of Pal~3, 
M3, and NGC~4147.

LB$^2$95 concluded that the measurement of proper motions for a
significant fraction of the globular cluster population would allow
more accurate association of objects following a stream in the Milky
Way halo.  However, the sample with measured proper motions consists
presently of only about one-fourth of all known Milky Way globular
clusters, and primarily those in the inner halo and the disk (i.e.,
those least likely to be stream members).  Nonetheless,Our ASPF
methodology will be increasingly useful as more proper motions become
available, and especially after the launch of the Space Interferometry
Mission (SIM), which will be able to measure proper motions of the
required accuracy (nearly a $\mu$arcsecond) for all Galactic satellites
and globular clusters.  We expect that the outermost globular clusters
($R_{apo} \gtrsim 40$ kpc), which have orbits that intersect the
spatial domain of those of the Milky Way DSGs, are the most likely to
have been accreted into the Milky Way's halo via tidal interactions
with their parent satellites.  Our GCPF analysis makes this connection
even more plausible.  As more proper motions are measured for outer
halo globular clusters (and DSGs) it is possible that
stronger evidence of accretion events in the outer halo will be found.

Although proper motions for outer halo globular clusters and Carina,
Leo I, Leo II, Fornax, and Sextans are of paramount importance, 
the associations made here can be followed up in other ways.
For example, detailed comparisons of the stellar populations of
dynamically associated globular clusters and their ``parent'' 
DSGs can be undertaken.  In addition, there should be stellar
streams associated with the alignment planes, which can be investigated
by searching for tidal tails of outer halo globular clusters as well
as applying the Great Circle Cell Counts technique \citep{kvj96}
to the streams' orbital planes.

\acknowledgements

We are especially grateful to Dana Dinescu for providing us with a copy
of her thesis in advance of publication.  We would also like to thank
Mike Irwin for providing us with a copy of a review article on dwarf
galaxy proper motions prior to its publication.  We thank Stephen Landy
for helpful discussions regarding the various methods of determining
$w(\theta)$, and also Andi Burkert and HongSheng Zhao for other helpful
conversations. We thank the original, anonymous referee for helpful
comments as well.  We would also like to thank Donald Lynden-Bell, the
second referee, for his careful reading and his many constructive
suggestions.  SRM and CP acknowledge support for this work from NSF
CAREER award AST-9702521, the David and Lucile Packard Foundation, and
a Cottrell Scholar Award from the Research Corporation.  KVJ was
supported in part by funds from the Institute for Advanced Study and
NASA LTSA grant NAG5-9064.

\appendix

\section{Random Data Generation}

The great circle orbital pole families were constructed for each object
using only the Cartesian $(X,Y,Z)$ Galactocentric radius vectors, and
therefore the random data generated to compare to this dataset were constructed
by randomizing only the distribution of radius vectors.   Rather than selecting
random points in $(X,Y,Z)$ space, we instead took the $(X,Y,Z)$ for each object
in our real sample and rotated it randomly.  This preserves the radial
distribution of the objects in the sample.

The Milky Way has a cylindrical symmetry (coordinates measured with
respect to the plane), but for the outer halo, a spherical symmetry.
In the case of the Milky Way globular clusters and DSGs, it was not
clear which was the more natural coordinate system to select.
Artificially generated data should be as realistic as possible; if
the real data has cylindrical symmetry, so should the artificial data
in order for the comparison to be fair.  For this reason, we created
two random datasets:  one where the objects were rotated at random with
respect to the Milky Way $Z$-axis and one where the objects were
rotated at random around all three axes.

For the ``cylindrical'' randomization, the object's $(X,Y)$ radius
components were multiplied by the simple rotation matrix (where $\theta$
was generated for each object in the sample with a standard random 
number generator such that $0 < \theta < 2\pi$):

\begin{equation}
\left[ \begin{array}{c} X^{\prime} \\ Y^{\prime} \end{array} \right] =
\left[ \begin{array}{lr} \cos\theta & -\sin\theta \\ \sin\theta & \cos\theta
\end{array} \right]
\left[ \begin{array}{c} X \\ Y \end{array} \right],
\end{equation}

\noindent For the ``spherical'' randomization, each object was 
rotated using the Eulerian rotation matrix (Marion \& Thornton 1988) (where
$\theta$, $\psi$, and $\phi$ were generated for each object in the sample
with a standard random number generator and $0 < \theta <
2\pi$, $0 < \phi < \pi$, and
$0 < \psi < \pi$):

\begin{equation}
\vec{X^{\prime}}=\lambda\vec{X},
\end{equation}

\noindent where $\vec{X} = (X,Y,Z)$ and,

\begin{equation}
\lambda = \left[ \begin{array}{ccc} \cos\psi\cos\phi - \cos\theta\sin\phi\sin\psi &
\cos\psi\sin\phi + \cos\theta\cos\phi\sin\psi & \sin\psi\sin\theta \\
-\sin\psi\cos\phi - \cos\theta\sin\phi\cos\psi & -\sin\psi\sin\phi + \cos\theta\cos\phi
i\cos\psi &
\cos\psi\sin\theta \\
\sin\theta\sin\phi & -\sin\theta\cos\phi & \cos\theta \\ \end{array} \right].
\end{equation}

When calculating the two point angular correlation function by
comparing data/data GCPF crossing point pairs to random/random GCPF
crossing point pairs, it was found that the amplitude, $w(\theta)$, was
independent of the randomization method used.  This suggests to us that
there may be no intrinsic cylindrical symmetry in the $R_{gc} > 8$ kpc
objects we studied, and thus spherical randomization creates artificial
data that is a fair comparison to the real data.

\clearpage
\pagestyle{empty}

\clearpage

\begin{deluxetable}{llrrl}
\small
\tablenum{1}
\tablecaption{Absolute Proper Motion Data for Globular Clusters and Satellite Galaxies}
\tablehead{
 & \colhead{Alternate} & \colhead{$\mu_{\alpha}\cos\delta$} &
\colhead{$\mu_{\delta}$} &  \\
\colhead{Name}   & \colhead{Name} &  \colhead{(mas/yr)} & \colhead{(mas/yr)} & \colhead{Reference} }
\startdata
NGC 104  & 47 Tuc  & $3.4\phn\pm1.7\phn\phn$ & $-1.9\phn\pm1.5\phn\phn$ & 
                     Cudworth \& Hanson 1993 \\
         &         & $6.43\pm2.10\phn$ & $-2.99\pm2.11\phn$ & 
                     Tucholke 1992a\tablenotemark{a} \\
         &         & $7.0\phn\pm1\phd\phn\phn\phn$ & $-5.3\phn\pm1\phd\phn\phn\phn$ & 
                     \citealt{oden} \\ 
NGC 288  & \nodata & $4.68\pm0.20\phn$ & $-5.25\pm0.19\phn$ & 
                     Guo 1995 \\ 
         &         & $4.67\pm0.42\phn$ & $-5.95\pm0.41\phn$ & 
                     Dinescu et al.\ 1999b\tablenotemark{b} \\ 
NGC 362  & \nodata & $4.43\pm1.02\phn$ & $-3.99\pm1.04\phn$ & 
                     Tucholke 1992b\tablenotemark{a} \\ 
         &         & $5.7\phn\pm1\phd\phn\phn\phn$ & $-1.1\phn\pm1\phd\phn\phn\phn$ & 
                     \citealt{oden} \\ 
NGC 1851 & \nodata & $1.28\pm0.68\phn$ & $2.39\pm0.65\phn$ & 
                     Dinescu et al.\ 1999b\tablenotemark{b} \\ 
NGC 1904 & M 79    & $2.12\pm0.64\phn$ & $-0.02\pm0.64\phn$ & 
                     Dinescu et al.\ 1999a \\ 
NGC 2298 & \nodata & $4.05\pm1.00\phn$ & $-1.72\pm0.98\phn$ & 
                     Dinescu et al.\ 1999a \\ 
Pal 3    & \nodata & $0.33\pm0.23\phn$ & $0.30\pm0.31\phn$ & 
                     Majewski \& Cudworth 1993 \\ 
NGC 4147 & \nodata & $-2.7\phn\pm1.3\phn\phn$ & $0.9\phn\pm1.3\phn\phn$ & 
                     Brosche et al.\ 1991 \\ 
         &         & $-1.0\phn\pm1\phd\phn\phn\phn$ & $-3.5\phn\pm1\phd\phn\phn\phn$ & 
                     \citealt{oden} \\ 
NGC 4590 & M 68    & $-3.76\pm0.66\phn$ & $1.79\pm0.62\phn$ & 
                     Dinescu et al.\ 1999a \\ 
NGC 5024 & M 53    & $0.5\phn\pm1\phd\phn\phn\phn$ & $-0.1\phn\pm1\phd\phn\phn\phn$ & 
                     \citealt{oden} \\ 
NGC 5139 & $\omega$ Cen & $-5.08\pm0.35\phn$ & $-3.57\pm0.34\phn$ & 
                     Dinescu et al.\ 1999a \\ 
NGC 5272 & M 3     & $-3.1\phn\pm0.2\phn\phn$ & $-2.3\phn\pm0.4\phn\phn$ & 
                     Scholz et al.\ 1993 \\ 
         &         & $-1.2\phn\pm2.5\phn\phn$ & $2.4\phn\pm3.0\phn\phn$  & 
                     Cudworth \& Hanson 1993 \\ 
         &         & $0.9\phn\pm1\phd\phn\phn\phn$ & $-2.2\phn\pm1\phd\phn\phn\phn$  & 
                     \citealt{oden} \\ 
         &         & $-1.2\phn\pm0.8\phn\phn$ & $-3.2\phn\pm0.8\phn\phn$ & 
                     Geffert 1998 \\ 
NGC 5466 & \nodata & $-5.4\phn\pm1.3\phn\phn$ & $0.6\phn\pm1.3\phn\phn$ & 
                     Brosche et al.\ 1991 \\ 
         &         & $-3.9\phn\pm1\phd\phn\phn\phn$ & $1.0\phn\pm1\phd\phn\phn\phn$ & 
                     \citealt{oden} \\ 
Pal 5    & \nodata & $-2.55\pm0.17\phn$ & $-1.93\pm0.17\phn$ & 
                     Cudworth et al.\ 2000 \\ 
         &         & $-1.0\phn\pm0.3\phn\phn$ & $-2.7\phn\pm0.4\phn\phn$ & 
                     Scholz et al.\ 1998 \\ 
         &         & $-2.44\pm0.17\phn$ & $-0.87\pm0.22\phn$ & 
                     Schweitzer et al.\ 1993 \\ 
NGC 5897 & \nodata & $-4.93\pm0.86\phn$ & $-2.33\pm0.84\phn$ & 
                     Dinescu et al.\ 1999a \\ 
NGC 5904\tablenotemark{c} & M 5 & $5.2\phn\pm1.7\phn\phn$ & $-14.2\phn\pm1.3\phn\phn$ & 
                     Cudworth \& Hanson 1993 \\ 
         &         & $6.7\phn\pm0.5\phn\phn$ & $-7.8\phn\pm0.4\phn\phn$ & 
                     Scholz et al.\ 1996 \\ 
         &         & $3.3\phn\pm1\phd\phn\phn\phn$ & $-10.1\phn\pm1\phd\phn\phn\phn$ & 
                     \citealt{oden} \\ 
NGC 6093 & M 80    & $-3.31\pm0.58\phn$ & $-7.20\pm0.67\phn$ & 
                     Dinescu et al.\ 1999a \\ 
NGC 6121 & M 4     & $-12.50\pm0.36\phn$ & $-19.93\pm0.49\phn$ & 
                     Dinescu et al.\ 1999a \\ 
         &         & $-11.6\phn\pm0.7\phn\phn$ & $-15.7\phn\pm0.7\phn\phn$ & 
                     Cudworth \& Hanson 1993 \\ 
NGC 6144 & \nodata & $-3.06\pm0.64\phn$ & $-5.11\pm0.72\phn$ & 
                     Dinescu et al.\ 1999a \\ 
NGC 6171 & M 107   & $-0.7\phn\pm0.9\phn\phn$ & $-3.1\phn\pm1.0\phn\phn$ & 
                     Cudworth \& Hanson 1993 \\ 
NGC 6205 & M 13    & $-0.9\phn\pm1.0\phn\phn$ & $5.5\phn\pm2.0\phn\phn$ & 
                     Cudworth \& Hanson 1993 \\ 
         &         & $-0.9\phn\pm1\phd\phn\phn\phn$ & $5.5\phn\pm1\phd\phn\phn\phn$ & 
                     \citealt{oden} \\ 
NGC 6218 & M 12    & $3.1\phn\pm0.6\phn\phn$ & $-7.5\phn\pm0.9\phn\phn$ & 
                     Scholz et al.\ 1996 \\ 
         &         & $1.6\phn\pm1.3\phn\phn$ & $-8.0\phn\pm1.3\phn\phn$ & 
                     Brosche et al.\ 1991 \\ 
         &         & $-0.8\phn\pm1\phd\phn\phn\phn$ & $-8.0\phn\pm1\phd\phn\phn\phn$ & 
                     \citealt{oden} \\ 
NGC 6254 & M 10    & $-6.0\phn\pm1\phd\phn\phn\phn$ & $-3.3\phn\pm1\phd\phn\phn\phn$ & 
                     \citealt{oden} \\ 
NGC 6341 & M 92    & $-4.4\phn\pm0.7\phn\phn$ & $1.1\phn\pm0.4\phn\phn$ & 
                     Scholz et al.\ 1994 \\ 
         &         & $-4.6\phn\pm1.1\phn\phn$ & $-0.6\phn\pm1.8\phn\phn$ & 
                     Cudworth \& Hanson 1993 \\ 
         &         & $-0.9\phn\pm1\phd\phn\phn\phn$ & $-1.5\phn\pm1\phd\phn\phn\phn$ & 
                     \citealt{oden} \\ 
         &         & $-4.4\phn\pm0.9\phn\phn$ & $-1.4\phn\pm0.9\phn\phn$ & 
                     Geffert 1998 \\ 
NGC 6362 & \nodata & $-3.09\pm0.46\phn$ & $-3.84\pm0.46\phn$ & 
                     Dinescu et al.\ 1999b\tablenotemark{b} \\ 
NGC 6397 & \nodata & $3.3\phn\pm0.5\phn\phn$ & $-15.2\phn\pm0.6\phn\phn$ & 
                     Cudworth \& Hanson 1993 \\ 
NGC 6522 & \nodata & $6.1\phn\pm0.2\phn\phn$ & $-1.8\phn\pm0.2\phn\phn$  & 
                     Terndrup et al.\ 1998 \\ 
NGC 6584 & \nodata & $-0.22\pm0.62\phn$ & $-5.97\pm0.64\phn$ & 
                     Dinescu et al.\ 1999b\tablenotemark{b} \\ 
NGC 6626 & M 28    & $0.3\phn\pm0.5\phn\phn$ & $-3.4\phn\pm0.9\phn\phn$ & 
                     Cudworth \& Hanson 1993 \\ 
NGC 6656 & M 22    & $8.6\phn\pm1.3\phn\phn$ & $-5.1\phn\pm1.5\phn\phn$ & 
                     Cudworth \& Hanson 1993 \\ 
NGC 6712 & \nodata & $4.2\phn\pm0.4\phn\phn$ & $-2.0\phn\pm0.4\phn\phn$ & 
                     Cudworth \& Hanson 1993 \\ 
NGC 6752 & \nodata & $-0.69\pm0.42\phn$ & $-2.85\pm0.45\phn$ & 
                     Dinescu et al.\ 1999b\tablenotemark{b} \\ 
NGC 6779 & M 56    & $0.3\phn\pm1\phd\phn\phn\phn$ & $1.4\phn\pm1\phd\phn\phn\phn$ & 
                     \citealt{oden} \\ 
NGC 6809 & M 55    & $-1.42\pm0.62\phn$ & $-10.25\pm0.64\phn$ & 
                     Dinescu et al.\ 1999a \\ 
NGC 6838 & M 71    & $-2.3\phn\pm0.8\phn\phn$ & $-5.1\phn\pm0.8\phn\phn$ & 
                     Cudworth \& Hanson 1993 \\ 
NGC 6934 & \nodata & $1.2\phn\pm1\phd\phn\phn\phn$ & $-5.1\phn\pm1\phd\phn\phn\phn$ & 
                     \citealt{oden} \\ 
NGC 7078\tablenotemark{c} & M 15 & $-0.3\phn\pm1.0\phn\phn$ & $-4.2\phn\pm1.0\phn\phn$ & 
                     Cudworth \& Hanson 1993 \\ 
         &         & $-1.0\phn\pm1.4\phn\phn$ & $-10.2\phn\pm1.4\phn\phn$ & 
                     Geffert et al.\ 1993 \\ 
         &         & $-0.1\phn\pm0.4\phn\phn$ & $0.2\phn\pm0.3\phn\phn$ & 
                     Scholz et al.\ 1996 \\ 
         &         & $-2.4\phn\pm1\phd\phn\phn\phn$ & $-8.3\phn\pm1\phd\phn\phn\phn$ & 
                     \citealt{oden} \\ 
NGC 7089 & M 2     & $5.5\phn\pm1.4\phn\phn$ & $-4.2\phn\pm1.4\phn\phn$ & 
                     Cudworth \& Hanson 1993 \\ 
         &         & $6.3\phn\pm1\phd\phn\phn\phn$ & $-5.7\phn\pm1\phd\phn\phn\phn$ & 
                     \citealt{oden} \\ 
NGC 7099 & M 30    & $1.42\pm0.69\phn$ & $-7.71\pm0.65\phn$ & 
                     Dinescu et al.\ 1999a \\ 
Pal 12   & \nodata & $-1.20\pm0.30\phn$ & $-4.21\pm0.29\phn$ & 
                     Dinescu et al.\ 2000 \\ 
Pal 13   & \nodata & $2.30\pm0.26\phn$ & $0.27\pm0.25\phn$ & Siegel et al.\ 2001 \\ 
LMC      & \nodata & $1.20\pm0.28\phn$ & $0.26\pm0.27\phn$ & 
                     Jones et al.\ 1994 \\ 
         &         & $1.3\phn\pm0.6\phn\phn$ & $1.1\phn\pm0.7\phn\phn$ & 
                     Kroupa et al.\ 1994 \\ 
         &         & $1.94\pm0.29\phn$ & $-0.14\pm0.36\phn$ & 
                     Kroupa \& Bastian 1997 \\ 
         &         & $1.60\pm0.29\phn$ & $0.19\pm0.37\phn$  &
                     van Leeuwen \& Evans 1998 \\
         &         & $1.7\phn\pm0.2\phn\phn$ & $2.8\phn\pm0.2\phn\phn$ & 
                     Anguita 1998 \\ 
SMC      & \nodata & $0.92\pm0.20\phn$ & $-0.69\pm0.20\phn$ & 
                     Irwin et al.\ 1996 \\ 
         &         & $0.5\phn\pm1.0\phn\phn$ & $-2.0\phn\pm1.4\phn\phn$ & 
                     Kroupa et al.\ 1994 \\ 
         &         & $1.23\pm0.84\phn$ & $-1.21\pm0.75\phn$ & 
                     Kroupa \& Bastian 1997 \\ 
         &         & $1.13\pm0.77\phn$ & $-1.17\pm0.66\phn$ &
                     van Leeuwen \& Evans 1998 \\
Ursa Minor & \nodata & $0.056\pm0.078$ & $0.078\pm0.099$ & 
                     Schweitzer et al.\ 1997 \\ 
           &         & $0.5\phn\pm0.8\phn\phn$ & $1.2\phn\pm0.5\phn\phn$ & 
                     Scholz \& Irwin 1994 \\ 
Draco    & \nodata & $0.6\phn\pm0.4\phn\phn$ & $1.1\phn\pm0.5\phn\phn$  & 
                     Scholz \& Irwin 1994 \\ 
Sculptor & \nodata & $0.72\pm0.22\phn$ & $-0.06\pm0.25\phn$ & 
                     Schweitzer et al.\ 1995 \\ 
Sagittarius & \nodata & $-2.65\pm0.08\phn$ & $-0.88\pm0.08\phn$  & 
                     Ibata et al.\ 2001b \\ 
\enddata

\tablenotetext{a}{Following Dinescu et al.\ 1999, we correct Tucholke's
relative proper motion with respect to SMC stars using the Kroupa \&
Bastian (1997) SMC proper motion determination.}

\tablenotetext{b}{Dinescu et al.\ 1999b revises the proper motions for
these objects from the values published in Dinescu et al.\ 1997.}

\tablenotetext{c}{Although the proper motions of Scholz et al.\ (1996)
have smaller errors, the discrepancies between their values and other
independent measurements are so large that we chose to use the Cudworth
\& Hanson (1993) values instead.}

\label{pmtab}

\end{deluxetable}

\clearpage

\begin{deluxetable}{lccrrrcrr}
\tabletypesize{\footnotesize}
\tablenum{2}
\tablecaption{Properties of Candidate Sagittarius Globular Clusters}
\tablehead{\colhead{ID} & \colhead{$E$\tablenotemark{a}} & 
\colhead{$L$\tablenotemark{a}} & 
\colhead{$R_{gc}$\tablenotemark{b}} &
\colhead{$R_{apo}$\tablenotemark{c}} & \colhead {$R_{peri}$\tablenotemark{c}} 
& \colhead{$[\textrm{Fe/H}]$\tablenotemark{b}} & \colhead{$M_V$\tablenotemark{b}} &
\colhead{$c$\tablenotemark{b}} \\
 & \colhead{($10^{4}$ km$^{2}$/sec$^{2}$)} & \colhead{($10^{2}$ kpc km/sec)} &
 kpc & kpc & kpc & & & \\ }

\startdata
Pal 5\tablenotemark{d} & $-7.4\pm0.8$ &  $23\pm7$  & 18.6 & 15.9 & 2.3 & -1.43 & 
-5.17 & 0.74 \\
Pal 5\tablenotemark{e} & $-7.3\pm0.5$ &  $21\pm10$  & 18.6 & 15.9 & 2.3 & -1.43 & 
-5.17 & 0.74 \\
M53 & $-4.4\pm2.5$ & $47\pm23$ & 18.8 & 36.0 & 15.5 & -1.99 & -8.77 & 1.78  \\
M5 & $-1.8\pm2.3$ & $14\pm5$ & 6.2 & 35.4 & 2.5 & -1.29 & -8.76 & 1.87 \\
NGC 5053 & \nodata & \nodata & 16.8 & \nodata & \nodata & -2.29 & -6.67 & 0.82 \\
NGC 6356 & \nodata & \nodata & 7.6 & \nodata & \nodata & -0.50 & -8.52 & 1.54 \\
\cline{1-9} 
M54 & \nodata & \nodata & 19.6 & \nodata & \nodata & -1.59 & -10.01 & 1.84 \\
Ter 7 &\nodata & \nodata & 16.0 & \nodata & \nodata & -0.58 & -5.05 & 1.08 \\
Arp 2 &\nodata & \nodata & 21.4 & \nodata & \nodata & -1.76 & -5.29 & 0.90 \\
Ter 8 & \nodata & \nodata & 19.1 & \nodata & \nodata & -2.00 & -5.05 & 0.60 \\
Pal 12 & $-6.0\pm2.1$ & $38\pm8$ & 15.9 & 29.0\tablenotemark{f} & 
16.0\tablenotemark{f} & -0.94 & -4.48 & 1.07 \\
\cline{1-9}
Sgr & $-4.4\pm0.6$ & $44\pm5$ & 24.0 & 54.0\tablenotemark{f} & 
14.0\tablenotemark{f} & -1.0\tablenotemark{g} &  \nodata &  \nodata\\
\enddata

\tablenotetext{a}{Integrated in the \citet{kvj95} potential}
\tablenotetext{b}{Taken from \citet{mwgc} compilation.}
\tablenotetext{c}{Taken from \citet{dd99b}.}
\tablenotetext{d}{\citet{kmc01p5} proper motion}
\tablenotetext{e}{\citet{scholz98} proper motion}
\tablenotetext{f}{Taken from \citet{dd00}.}
\tablenotetext{g}{Taken from \citet{mm98}.  Note that there is a dispersion
of $\sim$0.5 dex around this average value.}

\label{sgrgctab}
\end{deluxetable}

\clearpage

\begin{deluxetable}{ll|ll}
\tablewidth{265pt}
\tablenum{3}
\tablecaption{Zinn Horizontal Branch Types for Globular Clusters with $R_{gc} > 8$ kpc}
\tablehead{\multicolumn{2}{c}{Blue HB$+$Metal-Poor Type} &  
\multicolumn{2}{c}{Red HB Type} }
\startdata
NGC 288 & NGC 1904 & NGC 362 & NGC 1261\\ 
NGC 2298 & NGC 2419 & NGC 1851 & NGC 2808 \\ 
NGC 5024 & NGC 5053\tablenotemark{a} & NGC 3201 & NGC 4147 \\ 
NGC 5286 & NGC 5466\tablenotemark{a} & NGC 4590 & NGC 5272 \\ 
NGC 5694 & NGC 5824 & NGC 6229 & NGC 6864 \\ 
NGC 6101 & NGC 6205 & NGC 6934 & NGC 6981 \\ 
NGC 6341 & NGC 6426\tablenotemark{a} & NGC 7006 & Pal 3 \\ 
NGC 6715 & NGC 6779 & Pal 4 & Pal 5 \\ 
NGC 7078\tablenotemark{a} & NGC 7089 & Pal 12 & Pal 13 \\ 
NGC 7492 & Pal 1 & Pal 14 & Pyxis \\ 
IC 1257 &  Pal 15  & Arp 2 & Terzan 7 \\ 
         &          & AM 1 & Eridanus \\ 
         &          & IC 4499 & Rup 106 \\ 
\enddata

\tablenotetext{a}{Metal-Poor}

\label{hbtab}
\end{deluxetable}

\clearpage

\begin{deluxetable}{ccc}
\tablewidth{400pt}
\tablenum{4}
\tablecaption{Output of Centroid Clustering Algorithm for 6 Dwarf 
Galaxy Sample}
\tablehead{\colhead{Cluster Rank}  & {Pair of Objects}  & {Angular Separation}}
\startdata
1  &  SMC, U Min &  2.7$^{\circ}$ \\ 
2  &  LMC, Dra &  4.7$^{\circ}$ \\ 
3  &  SMC $\cup$ U Min, LMC $\cup$ Dra &  18.5$^{\circ}$ \\ 
4  &  Scl, SMC $\cup$ U Min $\cup$ LMC $\cup$ Dra 
& 60.9$^{\circ}$ \\ 
5  &  Sgr, SMC $\cup$ U Min $\cup$ LMC $\cup$ Dra $\cup$ Scl 
&  65.5$^{\circ}$ \\ 
\enddata

\label{dendtab}

\end{deluxetable}

\clearpage

\begin{deluxetable}{lrrr}
\tablenum{5}
\tablewidth{400pt}
\tablecaption{Derived Orbital Parameters for Objects with Grouped ASPFs}
\tablehead{ \colhead{Name} &  \colhead{$L_{tot}$} & \colhead{$L_{z}$}  
          & \colhead{$E$} \\ 
          & \colhead{($10^{2}$ kpc km/sec)}  & \colhead{($10^{2}$ kpc km/sec)}  
          &  \colhead{($10^{4}$ km$^{2}$/sec$^{2}$)}}
\startdata
Magellanic Group 1 & $dL_{tot} = 21$ & $dL_{z} = 2$ &  $dE = 1.1$ \\  \cline{1-4}
LMC            & $73\pm40$  & $-6\pm25$  &   $-2.1\pm0.9$  \\ 
NGC 7089 (M2)  & $20\pm10$   & $6\pm5$    &   $-5.9\pm2.1$  \\ 
NGC 6934       & $34\pm13$   & $3\pm8$   &   $-3.5\pm1.7$  \\ 
\\ 
Magellanic Group 2 & $dL_{tot} = 13$ & $dL_{z} = 8$ & $dE = 0.1$ \\   \cline{1-4}
Draco          & $430\pm173$ & $245\pm122$ &  $14.2\pm10.4$  \\ 
NGC 362        & $7\pm5$     & $5\pm3$     &  $-10.9\pm3.8$ \\ 
\\ 
Magellanic Group 3 & $dL_{tot} = 9$  & $dL_{z} = 5$ & $dE = 0.5$ \\  \cline{1-4}
SMC            & $70\pm39$   & $-37\pm20$  &  $-2.7\pm0.7$  \\ 
Ursa Minor     & $126\pm22$  & $-36\pm13$  &  $-0.7\pm0.4$  \\ 
Pal 3          & $341\pm126$ & $-108\pm74$ &  $5.1\pm3.0$   \\ 
NGC 5024 (M53) & $47\pm23$   & $-13\pm5$   &  $-4.4\pm2.5$  \\ 
\\ 
Sagittarius Group\tablenotemark{a} 
               & $dL_{tot} = 2 - 4$ & $dL_{z} = 0.07 - 0.14$ & $dE = 0.1 - 0.3$ \\  \cline{1-4}
Sagittarius    & $44\pm5$    & $2\pm3$     &  $-4.4\pm0.6$  \\ 
NGC 5466       & $40\pm24$   & $9\pm6$     &  $-1.3\pm2.7$  \\ 
\\ 
Sculptor Group & $dL_{tot} = 3$ & $dL_{z} = 0.4$  & $dE = 0.1$ \\  \cline{1-4}
Sculptor       & $159\pm100$  & $-20\pm9$   &   $0.3\pm2.1$  \\ 
NGC 6584       & $6\pm4$      & $-3\pm2$    &   $-9.8\pm0.9$ \\ 
Pal 5\tablenotemark{b}  & $21\pm10$ & $-4\pm4$ & $-7.3\pm0.5$   \\ 
NGC 5904 (M5)  & $14\pm5$ & $-1\pm2$ & $-1.8\pm2.3 $ \\ 
NGC 6144       & $5\pm2$ & $1\pm1$ & $-13.8\pm0.5$ \\ 
\enddata

\tablenotetext{a}{The values listed for $dL_{tot}$, $dL_{z}$, and
$dE$ are for a range of $M_{Sgr}$ from $10^{7} - 10^{8} M_{\sun}$.}
\tablenotetext{b}{The values listed in the table are for the ASPF of Pal 5
calculated using the Scholz et al.\ (1998) proper motion.  The ASPF for
Pal 5 from the Cudworth et al.\ (2000) proper motion is in a different part of 
the sky and has a different value of $L_{z}$ and $E$.}

\label{orbtab}

\end{deluxetable}

\end{document}